\documentclass[11pt]{article}
\usepackage{geometry}

\usepackage{graphics,graphicx,color,float, hyperref}
\usepackage{epsfig,bbm}
\usepackage{bm}
\usepackage{mathtools}
\usepackage{amsmath}
\usepackage{amssymb}
\usepackage{amsfonts}
\usepackage{amsthm}
\usepackage {times}
\usepackage{layout}
\usepackage{setspace}
\usepackage{tikz-cd}
\allowdisplaybreaks[3]

\newtheorem{fact}{Fact}

\newcommand{\beq}{\begin{equation}}
\newcommand{\enq}{\end{equation}}
\newcommand{\bel}{\begin{lemma}}
\newcommand{\enl}{\end{lemma}}
\newcommand{\bet}{\begin{theorem}}
\newcommand{\ent}{\end{theorem}}

\newcommand{\tr}{\mathrm{Tr}}

\newcommand{\Tr}{\mathrm{Tr}}
\newcommand{\ketbra}[1]{|#1\rangle\langle#1|}

\newcommand{\eps}{\varepsilon}

\newcommand{\dzeroseps}[1]{\ensuremath{\mathrm{D}_{\mathrm{H}}^{#1}}}

\newcommand*{\cH}{\mathcal{H}}

\newcommand*{\cD}{\mathcal{D}}

\newcommand*{\cN}{\mathcal{N}}

\newcommand*{\cE}{\mathcal{E}}

\newcommand*{\cU}{\mathcal{U}}

\newcommand{\suppress}[1]{}

\newcommand{\defeq}{\ensuremath{ \stackrel{\mathrm{def}}{=} }}
\newcommand{\F}{\mathrm{F}}
\newcommand{\Pur}{\mathrm{P}}
\newcommand {\br} [1] {\ensuremath{ \left( #1 \right) }}

\newcommand {\minusspace} {\: \! \!}

\newcommand {\fn} [2] {\ensuremath{ #1 \minusspace \br{ #2 } }}
\newcommand {\ball} [2] {\fn{\mathcal{B}^{#1}}{#2}}
\newcommand {\relent} [2] {\fn{\mathrm{D}}{#1 \middle\| #2}}
\newcommand {\dmax} [2] {\fn{\mathrm{D}_{\max}}{#1 \middle\| #2}}

\newcommand {\dheps} [3] {\ensuremath{\mathrm{D}_{\mathrm{H}}^{#3}\left(#1 \| #2\right)}}
\newcommand {\id} {\ensuremath{\mathrm{I}}}

\newcommand{\bra}[1]{\langle #1|}
\newcommand{\ket}[1]{|#1 \rangle}

\mathchardef\mhyphen="2D

\makeatletter
\newcommand*{\rom}[1]{\expandafter\@slowromancap\romannumeral #1@}
\makeatother

\mathchardef\mhyphen="2D

\newtheorem{definition}{Definition}
\newtheorem{claim}{Claim}

\newtheorem{theorem}{Theorem}
\newtheorem{lemma}{Lemma}
\newtheorem{corollary}{Corollary}

\begin {document}
\title{On the near-optimality of one-shot classical communication over quantum channels}
\author{
Anurag Anshu\footnote{Center for Quantum Technologies, National University of Singapore, Singapore. \texttt{a0109169@u.nus.edu}} \qquad
Rahul Jain\footnote{Center for Quantum Technologies, National University of Singapore and MajuLab, UMI 3654, 
Singapore. \texttt{rahul@comp.nus.edu.sg}} \qquad 
Naqueeb Ahmad Warsi\footnote{Center for Quantum Technologies, National University of Singapore and School of Physical and Mathematical Sciences, Nanyang Technological University, Singapore and IIITD, Delhi. \texttt{warsi.naqueeb@gmail.com}} 
}
\date{}
\maketitle

\abstract{We study the problem of transmission of classical messages through a quantum channel in several network scenarios in the one-shot setting. We consider both the entanglement assisted and unassisted cases for the point to point quantum channel, quantum multiple-access channel, quantum channel with state and the quantum broadcast channel. We show that it is possible to near-optimally characterize the amount of communication that can be transmitted in these scenarios, using the position-based decoding strategy introduced in a prior work \cite{AnshuJW17CC}. In the process, we provide a short and elementary proof of the converse for entanglement-assisted quantum channel coding in terms of the quantum hypothesis testing divergence (obtained earlier in \cite{MatthewsW14}). Our proof has the additional utility that it naturally extends to various network scenarios mentioned above. Furthermore, none of our achievability results require a \textit{simultaneous decoding} strategy, existence of which is an important open question in quantum Shannon theory. 
}

\section{Introduction}
\label{intro}

Understanding the limits of communication through various models of channels is a central aspect of classical information theory. Some landmark results in this direction are the models of point to point channel \cite{Shannon}, multiple access channel \cite{Ahlswede73, Liao72}, channel with a state \cite{GelfandP80} and broadcast channel \cite{Marton79}. The diversity of scenarios in which information theory can be applied has led to various settings in which the problem of channel coding is studied. Below we discuss two settings relevant to this work. 
\begin{itemize}
\item \textbf{Asymptotic and i.i.d. setting:} Here, the senders are allowed to use the channel multiple times in a memoryless fashion and the goal is to obtain bounds on the rate of transmission for an arbitrarily large number of channel uses, as the error is made to go to zero. It is highly desirable that the resulting bounds are single letter, that is, they do not require unbounded optimization in their computation. Without this restriction, it would be possible to obtain tight characterization of the capacity of all of the aforementioned channel settings \cite[Section 4.3]{GamalK12}.  

\item \textbf{One-shot setting:} Here, the senders are allowed to use the channel only once, which can arise in many practical scenarios. It is desirable to obtain bounds on the amount of communication which are near optimal. That is, a communication cost (or cost region for multiple messages) $R(\eps)$ may be obtained which is a converse cost for any protocol that makes an error $\eps$ (in terms of the probability of incorrectly decoding the messages) and there exists a protocol that achieves the cost $R(\eps')$ (where $\eps'$ is of the order of $\eps$) up to some additive factors. 
\end{itemize} 

Quantum information theory generalizes the models of classical channels in various ways, by introducing channels that can take quantum inputs and produce quantum outputs or by allowing new resources such as quantum entanglement. Several works have studied the problem of transmission of quantum information through a point to point quantum channel in the asymptotic and i.i.d. setting, both in the entanglement assisted case \cite{BennettSST02} and the entanglement unassisted cases (\cite{Holevo98, SchuW97} for the transmission of classical information and \cite{lloyd97, Shor02, Devetak05private} for the transmission of quantum information). In the entanglement assisted case, the transmission of classical information is equivalent to the transmission of quantum information up to a factor of $2$, due to the duality between quantum teleportation \cite{Teleportation93} and super-dense coding \cite{bennett92}. In the entanglement unassisted case, the duality is lost and we have two different aforementioned scenarios for the transmission of classical information and quantum information. In this work, we shall focus on the transmission of classical information for both of the entanglement assisted and unassisted cases. 

Several quantum network scenarios have also been studied in the asymptotic and i.i.d. setting, such as the quantum multiple access channel \cite{Winter01, HsiehDW08, YardHD08, FawziHSSW12, XuW13}, the quantum broadcast channel \cite{SaakianA98, DupuisHL10, YardHD11, SavovW15} and quantum channel with state \cite{Dupuis09, Frederic10}. In most of these cases (both the entanglement assisted and unassisted), a single letter characterization is not known. Some exceptions, where a single letter characterization is known, are the entanglement assisted point to point quantum channel \cite{BennettSST02}, the classical-quantum multiple access channel \cite{Winter01} and the entanglement assisted quantum channel with state \cite{Dupuis09, Frederic10}.   

Classical communication over the point to point channel has been studied in several works in the one-shot setting \cite{Frederic10, BuscemiD10, WangR12, DattaH13, DattaTW2016, AnshuJW17CC}. These results have been extended to the quantum network scenarios in the works \cite{Dupuis09, DupuisHL10, Frederic10, AnshuJW17CC}. However, in all the cases except for the point to point channel (both entanglement assisted \cite{AnshuJW17CC} and entanglement unassisted \cite{WangR12}), a near-optimal one-shot characterization is not known. An interesting variant, where the communicating parties are equipped with arbitrary non-local correlations, has also been considered in the works \cite{WangFT17, XieWD17}, providing improvements to the entanglement assisted case (which is a weaker non-local resource).

In this work, we provide a near-optimal one-shot characterization for many quantum network scenarios, using the position-based decoding strategy introduced in \cite{AnshuJW17CC}. In contrast with \cite{AnshuJW17CC}, we do not require the convex-split technique \cite{AnshuDJ14} for our achievability results. Our results, as obtained in Sections \ref{sec:entqcode} and \ref{sec:unassistqcode}, are summarized below. 
\begin{itemize}
\item \textbf{Point to point quantum channel:} A converse bound for the entanglement assisted case has been given in \cite{MatthewsW14} and a nearly matching achievability result has been obtained recently in \cite{AnshuJW17CC}. We provide a short proof of the converse in \cite{MatthewsW14}. This also considerable simplifies an alternative proof given in \cite[arXiv version 2]{AnshuJW17CC}, which was inspired by the analogous asymptotic and i.i.d. result \cite[Section 21.5]{Wilde17} and used a one-shot analogue of the chain rule for the conditional quantum mutual information. We are able to avoid the use of any such chain rule in our converse proof, by considering the quantum hypothesis testing divergence between appropriate quantum states. Our proof technique has the utility that it easily extends to various network scenarios. As an application, we recover the one-way case of \cite[Theorem 1.1]{NayakS02}.

For the entanglement unassisted case, we provide a similar characterization to that given in \cite{WangR12}. Our characterization has the property that it is of similar form for other entanglement unassisted network scenarios.

\item \textbf{Quantum channel with state:} We provide a near optimal characterization for this channel in the one-shot setting, with a tight dependence on the error of decoding. The optimization involved in our bound is comparable to the optimization involved in earlier known results \cite{Dupuis09, Frederic10, AnshuJW17CC}. It is not clear if our bound attains a single letter expression in the asymptotic and i.i.d. setting, in contrast with the asymptotic and i.i.d. form of the bounds given in \cite{Dupuis09, Frederic10, AnshuJW17CC}. On the other hand, we show as a corollary that the achievability bound given in \cite{AnshuJW17CC} for the quantum channel with state is near optimal in the one-shot setting. Same feature is not known for the one-shot achievability bounds in \cite{Dupuis09, Frederic10}. 

We also provide near-optimal bounds for this channel for the entanglement unassisted case, with the property that the registers involved in our bounds have dimension comparable to that of the input and output registers of the channel.

\item \textbf{Quantum broadcast channel:} In a similar fashion to the quantum channel with state, we provide a near optimal characterization for this channel in the one-shot setting (discussing the case of one sender and two receivers, as the results similarly extend to more than two receivers), with a tight dependence on the error of decoding. The optimization involved in our bound is comparable to the optimization involved in earlier known results \cite{DupuisHL10, Frederic10, AnshuJW17CC}.  We note that the asymptotic and i.i.d. analogue of our converse result is implicit in \cite[Theorem 3]{DupuisHL10}. It is not clear if our bound attains a single letter expression in the asymptotic and i.i.d. setting, which is also the case for the asymptotic and i.i.d. form of the bounds given in \cite{DupuisHL10, Frederic10, AnshuJW17CC}. On the other hand, we show as a corollary that the achievability bound given in \cite{AnshuJW17CC} for the quantum channel with state is near optimal in the one-shot setting. Same feature is not known for the one-shot achievability bound in \cite{DupuisHL10, Frederic10}.

We also provide near-optimal bounds for this channel for the entanglement unassisted case, with the property that the registers involved in our bounds have dimension comparable to that of the input and output registers of the channel.

\item \textbf{Quantum multiple access channel:}   We provide a new converse bound for the multiple access channel with two senders and one receiver (which can easily be extended to the case of more than two senders). We show how to achieve this bound in two different ways (both of which can easily be extended to the case of more than two senders). The first way uses the pretty good measurement technique of Hayashi and Nagaoka \cite{HayashiN03} and has a tight dependence on the error of decoding one of the messages (at the cost of quadratic loss on the error of decoding the other message). The second way uses the sequential decoding strategy of Sen \cite{Sen12} (with the quantitatively improved version of \cite{Gao15}; see also the related works \cite{GioLM12, Wilde13} and the recent improvement \cite{OMW18}) and has a tight dependence on the error of decoding both messages up to multiplicative constants. As far as we know, this is a first instance where the sequentially decoding strategy gives a better dependence on the overall error of decoding the messages in comparison to the pretty good measurement. Furthermore, our achievability results \textit{do not} need a simultaneous decoding strategy \cite{FawziHSSW12, DrescherF13, QiWW17}. It is not clear if this bound leads to a single letter characterization in the asymptotic and i.i.d. setting, a situation similar to the other known bound for the entanglement assisted quantum multiple access channel in the asymptotic and i.i.d. setting \cite{HsiehDW08}.  

We also provide near-optimal bounds for the entanglement unassisted case, with the property that the registers involved in our bounds have dimension comparable to that of the input and output registers of the channel.

\end{itemize}

\section{Preliminaries}
\label{sec:prelim}
In this section we set our notations, make the definitions and state the facts that we will need later for our proofs. 

Consider a finite dimensional Hilbert space $\cH$ endowed with an inner product $\langle \cdot, \cdot \rangle$. The $\ell_1$ norm of an operator $X$ on $\cH$ is $\| X\|_1:=\Tr\sqrt{X^{\dagger}X}$ and $\ell_2$ norm is $\| X\|_2:=\sqrt{\Tr XX^{\dagger}}$. For hermitian operators $X, X'$, the notation $X\preceq X'$ implies that $X' - X$ is a positive semi-definite operator. A quantum state (or a density matrix or a state) is a positive semi-definite matrix on $\cH$ with trace equal to $1$. It is called {\em pure} if and only if its rank is $1$. A sub-normalized state is a positive semi-definite matrix on $\cH$ with trace less than or equal to $1$. Let $\ket{\psi}$ be a unit vector on $\cH$, that is $\langle \psi,\psi \rangle=1$.  With some abuse of notation, we use $\psi$ to represent the state and also the density matrix $\ketbra{\psi}$, associated with $\ket{\psi}$.  Given a quantum state $\rho$ on $\cH$, the {\em support of $\rho$}, called $\text{supp}(\rho)$ is the subspace of $\cH$ spanned by all eigenvectors of $\rho$ with non-zero eigenvalues. For quantum states $\rho,\sigma$ on $\cH$, the notation $\text{supp}(\rho)\subseteq \text{supp}(\sigma)$ means that the support of $\rho$ is contained in the support of $\sigma$.

A {\em quantum register} $A$ is associated with some Hilbert space $\cH_A$. Define $|A| := \dim(\cH_A)$. Let $\mathcal{L}(A)$ represent the set of all linear operators acting on the set of quantum states acting on the Hilbert space $\cH_A$. We denote by $\mathcal{D}(A)$, the set of quantum states on the Hilbert space $\cH_A$. State $\rho$ with subscript $A$ indicates $\rho_A \in \mathcal{D}(A)$. If two registers $A,B$ are associated with the same Hilbert space, we shall represent the relation by $A\equiv B$.  Composition of two registers $A$ and $B$, denoted $AB$, is associated with Hilbert space $\cH_A \otimes \cH_B$.  For two quantum states $\rho\in \mathcal{D}(A)$ and $\sigma\in \mathcal{D}(B)$, $\rho\otimes\sigma \in \mathcal{D}(AB)$ represents the tensor product (Kronecker product) of $\rho$ and $\sigma$. The identity operator on $\cH_A$ (and associated register $A$) is denoted $\id_A$. 

Let $\rho_{AB} \in \mathcal{D}(AB)$. We define
\[ \rho_{B} := \Tr_{A}(\rho_{AB}) := \sum_i (\bra{i} \otimes \id_{B}) \rho_{AB} (\ket{i} \otimes \id_{B}) , \]
where $\{\ket{i}\}_i$ is an orthonormal basis for the Hilbert space $\cH_A$.
The state $\rho_B\in \mathcal{D}(B)$ is referred to as the marginal state of $\rho_{AB}$. Unless otherwise stated, a missing register from subscript in a state will represent partial trace over that register. Given a $\rho_A\in\mathcal{D}(A)$, a {\em purification} of $\rho_A$ is a pure state $\rho_{AB}\in \mathcal{D}(AB)$ such that $\Tr_{B}(\rho_{AB})=\rho_A$. Purification of a quantum state is not unique.

A quantum {map} $\cE: \mathcal{L}(A)\rightarrow \mathcal{L}(B)$ is a completely positive and trace preserving (CPTP) linear map (mapping states in $\mathcal{D}(A)$ to states in $\mathcal{D}(B)$). A {\em unitary} operator $U_A:\cH_A \rightarrow \cH_A$ is such that $U_A^{\dagger}U_A = U_A U_A^{\dagger} = \id_A$. An {\em isometry}  $V:\cH_A \rightarrow \cH_B$ is such that $V^{\dagger}V = \id_A$ and $VV^{\dagger} = \Pi_B$, where $\Pi_B$ is a projection on $\cH_B$. The set of all unitary operations on register $A$ is  denoted by $\mathcal{U}(A)$.

We shall consider the following information theoretic quantities. Let $\varepsilon \in (0,1)$. 
\begin{enumerate}
\item {\bf Fidelity} (\cite{Josza94}, see also \cite{uhlmann76}). For $\rho_A,\sigma_A \in \mathcal{D}(A)$, $$\F(\rho_A,\sigma_A)\defeq\|\sqrt{\rho_A}\sqrt{\sigma_A}\|_1.$$ 
\item {\bf Purified distance} (\cite{GilchristLN05}). For $\rho_A,\sigma_A \in \mathcal{D}(A)$, $$\Pur(\rho_A,\sigma_A) = \sqrt{1-\F^2(\rho_A,\sigma_A)}.$$
\item {\bf $\varepsilon$-ball}. For $\rho_A\in \mathcal{D}(A)$, $$\ball{\eps}{\rho_A} \defeq \{\rho'_A\in \mathcal{D}(A)|~\Pur(\rho_A,\rho'_A) \leq \varepsilon\}. $$ 

\item {\bf Relative entropy} (\cite{umegaki1954}). For $\rho_A,\sigma_A\in \mathcal{D}(A)$ such that $\text{supp}(\rho_A) \subseteq \text{supp}(\sigma_A)$, $$\relent{\rho_A}{\sigma_A} \defeq \Tr(\rho_A\log\rho_A) - \Tr(\rho_A\log\sigma_A) .$$ 
\item {\bf Smooth quantum hypothesis testing divergence} (\cite{BuscemiD10}, see also \cite{HayashiN03}).  For $\rho_A,\sigma_A\in \mathcal{D}(A)$ and $\eps \in (0,1)$, $$ \mathrm{D}_{\mathrm{H}}^{\eps}(\rho_A\| \sigma_A)  \defeq  \sup_{0\preceq \Lambda \preceq \id_A, \Tr(\Lambda\rho_A)\geq 1-\eps}\log\left(\frac{1}{\Tr(\Lambda\sigma_A)}\right).$$  
\item {\bf Max-relative entropy} (\cite{Datta09}). For $\rho_A,\sigma_A\in \mathcal{D}(A)$ such that $\text{supp}(\rho_A) \subset \text{supp}(\sigma_A)$, $$ \dmax{\rho_A}{\sigma_A}  \defeq  \inf \{ \lambda \in \mathbb{R} : \rho_A \preceq 2^{\lambda} \sigma_A  \}  .$$  
\end{enumerate}

We will use the following facts. 
\begin{fact}[Triangle inequality for purified distance,~\cite{GilchristLN05, Tomamichel12}]
\label{fact:trianglepurified}
For states $\rho_A, \sigma_A, \tau_A\in \mathcal{D}(A)$,
$$\Pur(\rho_A,\sigma_A) \leq \Pur(\rho_A,\tau_A)  + \Pur(\tau_A,\sigma_A) . $$ 
\end{fact}

\begin{fact}[Monotonicity under quantum operations, \cite{barnum96},\cite{lindblad75}]
	\label{fact:monotonequantumoperation}
For quantum states $\rho$, $\sigma \in \mathcal{D}(A)$, and quantum operation $\cE(\cdot):\mathcal{L}(A)\rightarrow \mathcal{L}(B)$, it holds that
\begin{align*}
	\F(\cE(\rho),\cE(\sigma)) \geq \F(\rho,\sigma) \quad \mbox{and} \quad \dzeroseps{\eps}(\rho\|\sigma) \geq \dzeroseps{\eps}(\cE(\rho)\|\cE(\sigma)).
\end{align*}
\end{fact}

\begin{fact}[\cite{AnshuJW17CC}]
\label{closestatesmeasurement}
Let $\rho,\sigma$ be quantum states and $0\leq \Pi\leq \id$ be an operator. Then $|\sqrt{\Tr(\Pi\sigma)} - \sqrt{\Tr(\Pi\rho)}|\leq \Pur(\rho, \sigma)$.
\end{fact}

\begin{fact}[Gentle measurement lemma,\cite{Winter:1999,Ogawa:2002}]
\label{gentlelemma}
Let $\rho$ be a quantum state and $0<A<\id$ be an operator. Then 
$$\F(\rho, \frac{A\rho A}{\Tr(A^2\rho)})\geq \sqrt{\Tr(A^2\rho)}.$$
\end{fact}

Following fact is analogous to the gentle measurement lemma (Fact \ref{gentlelemma}).
\begin{fact}
\label{gentlepovm}
Consider a quantum state $\ket{\rho}$ and a measurement $\{A_i\}_i$ such that $A_i \succeq 0$. Let $\rho'\defeq \sum_i A_i\ketbra{\rho} A_i$. Then $\F^2(\rho, \rho')= \sum_i \Tr(A_i\rho)^2 \geq \sum_i \Tr(A^2_i\rho)^2$. 
\end{fact}

\begin{fact}[Hayashi-Nagaoka inequality, \cite{HayashiN03}]
\label{haynag}
Let $0\preceq S\preceq \id,T$ be positive semi-definite operators and $c > 0$. Then 
$$\id - (S+T)^{-\frac{1}{2}}S(S+T)^{-\frac{1}{2}}\preceq (1+c)(\id-S) + (2 + c + \frac{1}{c}) T.$$
\end{fact}

\begin{fact}[\cite{WangR12}]
\label{dhandd}
Let $\eps\in (0,1)$ and $\rho_A, \sigma_A\in \mathcal{D}(A)$. It holds that 
$$\mathrm{D}_{\mathrm{H}}^{\eps}(\rho_A\| \sigma_A) \leq \frac{\relent{\rho_A}{\sigma_A}}{1-\eps}.$$
\end{fact}

\begin{fact}[Sequential measurement, \cite{Sen12, Gao15}]
\label{noncommutativebound}
Let $\rho$ be a quantum state and $\Pi_1, \Pi_2, \ldots \Pi_k$ be projectors. Let $\Pi'_i \defeq \id - \Pi_i$. Then 
$$\Tr(\Pi'_k\Pi'_{k-1}\ldots\Pi'_1\rho\Pi'_1\Pi'_2\ldots\Pi'_k)\geq 1- 4\sum_i\Tr(\Pi_i\rho).$$
\end{fact}

\begin{fact}
\label{fact:dhallstates}
Let $\eps\in (0,1)$. Let $\rho_{MM'}$ be a quantum state such that $\rho_M = \frac{\id_M}{|M|}$ and $\Tr(\sum_m\ketbra{m}_M\otimes \ketbra{m}_{M'}\rho_{MM'})\geq 1-\eps$. Then for any quantum state $\sigma_{M'}$, 
$$\dheps{\rho_{MM'}}{\rho_M\otimes \sigma_{M'}}{\eps} \geq \log|M|.$$
\end{fact}
\begin{proof}
Setting $\Lambda_{MM'}= \sum_m\ketbra{m}_M\otimes \ketbra{m}_{M'}$, consider $$\Tr(\Lambda_{MM'}\rho_M\otimes\sigma_{M'}) = \sum_m\bra{m}_M\rho_M\ket{m}_M\cdot \bra{m}_{M'}\sigma_{M'}\ket{m}_{M'} = \frac{1}{|M|}\sum_m\bra{m}_{M'}\sigma_{M'}\ket{m}_{M'} = \frac{1}{|M|}.$$ Further, $\Tr(\Lambda_{MM'}\rho_{MM'})\geq 1-\eps$. The bound now follows from the definition of  $\dheps{\rho_{MM'}}{\rho_M\otimes \sigma_{M'}}{\eps}$.
\end{proof}

\begin{fact} [Neumark's theorem, \cite{Watrouslecturenote}] \label{Neumark}For any POVM $\left\{M_i\right\}_{i \in \mathcal{I}}$ acting on a system $S,$ there exists a unitary $U_{SP}$ and an orthonormal basis $\left\{\ket{i}_P\right\}_{i \in \mathcal{I}}$ such that for all quantum states $\rho_S$, we have
$$\tr \left[U^\dagger_{SP} \left(\mathbb{I}_S \otimes \ket{i}\bra{i}_P\right)U_{SP}\left(\rho_S \otimes \ket{0}\bra{0}_P\right)\right] = \tr \left[ M_i \rho_S\right].$$
\end{fact}

\begin{fact} 
\label{fact:puredheps}
Let $\eps\in (0,1)$ and $\ketbra{\rho}_A, \sigma_A$ be quantum states. Then 
$$\dheps{\ketbra{\rho}_A}{\sigma_A}{\eps} =  \sup_{0\preceq \Lambda \preceq \id_A: \mathrm{rk}(\Lambda) =1, \Tr(\Lambda\ketbra{\rho}_A)\geq 1-\eps}\log\left(\frac{1}{\Tr(\Lambda\sigma_A)}\right),$$ where $\mathrm{rk}(\Lambda)$ is the rank of the operator $\Lambda$.
\end{fact}
\begin{proof}
We apply Neumark's theorem (Fact \ref{Neumark}) to rewrite the smooth quantum hypothesis divergence as
\begin{equation}
\label{eq:pureproj}
\sup_{\Pi: \Pi^2=\Pi,  \Tr(\Pi\ketbra{\rho}_A\otimes \ketbra{0}_P)\geq 1-\eps}\log\left(\frac{1}{\Tr(\Pi\sigma_A \otimes \ketbra{0}_P)}\right).
\end{equation}
Fix a $\Pi$ such that $ \Tr(\Pi\ketbra{\rho}_A\otimes \ketbra{0}_P)\geq 1-\eps$, define $\ketbra{\phi} \defeq \frac{\Pi\ketbra{\rho}_A\otimes \ketbra{0}_P\Pi}{\bra{\rho}_A\bra{0}_P\Pi\ket{0}_P\ket{\rho}_A}$. Then
$$\Tr(\ketbra{\phi}\cdot\ketbra{\rho}_A\otimes \ketbra{0}_P) =  \frac{\bra{\rho}_A\bra{0}_P\Pi\ket{\rho}_A\ket{0}_P\cdot\bra{\rho}_A\bra{0}_P\Pi\ket{0}_P\ket{\rho}_A}{\bra{\rho}_A\bra{0}_P\Pi\ket{0}_P\ket{\rho}_A} = \bra{\rho}_A\bra{0}_P\Pi\ket{0}_P\ket{\rho}_A \geq 1-\eps.$$
Further, since $\Pi\ket{\phi} = \ket{\phi}$, we have $\ketbra{\phi}\preceq \Pi$. Thus, $\Tr(\ketbra{\phi}\cdot \sigma_A\otimes \ketbra{0}_P) \leq \Tr(\Pi\sigma_A\otimes \ketbra{0}_P)$. Thus, the projector achieving the supremum in Equation \ref{eq:pureproj} has rank $1$. The proof now follows by defining $\Lambda \defeq \bra{0}_P\ketbra{\phi}\ket{0}_P$, which satisfies $\mathrm{rk}(\Lambda)=1$.
\end{proof}

\section{Entanglement assisted quantum coding}
\label{sec:entqcode}

\subsection{Point to point quantum channel}

Alice wants to communicate a classical message $M$ chosen from $[2^R]$ to Bob over a quantum channel such that Bob is able to decode the correct message with probability at least $1-\eps$ , for all message $m$. To accomplish this task Alice and Bob also share entanglement between them.  Let the input to Alice be given in a register $M$. We now make the following definition.
\begin{definition}
Let $\ket{\theta}_{E_AE_B}$ be the shared entanglement between Alice ($E_A$) and Bob ($E_B$). An $(R, \eps )$-entanglement assisted code for the quantum channel $\cN_{A \to B}$ consists of 
\begin{itemize}
\item An encoding map $\cU : ME_A \rightarrow A$ for Alice.  
\item A decoding operation $\cD : BE_B\rightarrow M'$ for Bob, with $M'\equiv M$ being the output register such that for all $m \in [2^R]$,
\beq
\Pr(M'\neq m|M=m) \leq \eps. \nonumber
\enq
\end{itemize}
\end{definition}

The following converse was shown in \cite{MatthewsW14}. We provide a simpler proof with the utility that it can be easily extended to complex network scenarios.

\begin{theorem}
\label{entptopconverse}
Fix a quantum channel $\cN_{A\to B}$ and $\eps\in (0,1)$. For any $(R, \eps)$-entanglement assisted code for this quantum channel, it holds that 
$$R \leq \max_{\ket{\psi}_{AB'}} \min_{\sigma_B}\dzeroseps{\eps}(\cN_{A\to B}(\psi_{AB'})\|\sigma_B\otimes \psi_{B'}).$$ 
\end{theorem}
\begin{proof}
\suppress{
Let $\ket{\Phi}_{AB'}\otimes \ket{\Phi}_{E_AE_B}$ be the unnormalized maximally entangled state. Observe that 
$\ket{\theta}_{AE_AB'E_B} = (\id_{AE_A}\otimes \theta^{\frac{1}{2}}_{B'E_B})\ket{\Phi}_{AB'}\otimes \ket{\Phi}_{E_AE_B}$. After transmission through the channel for a given $m$, the quantum state on Bob's registers is 
\begin{eqnarray*}
&&\cN_{A \to B}\left((\id_A\otimes\theta^{\frac{1}{2}}_{B'E_B}U^T(m)_{B'E_B})\ketbra{\Phi}_{AB'}\otimes \Phi_{E_B}(\id_A\otimes U^{T\dagger}(m)_{B'E_B}\theta^{\frac{1}{2}}_{B'E_B})\right)\\
&& = (\id_A\otimes\theta^{\frac{1}{2}}_{B'E_B}U^T(m)_{B'E_B}\theta^{-\frac{1}{2}}_{B'})\cN_{A \to B}(\ketbra{\theta}_{AB'})\otimes \Phi_{E_B}(\id_A\otimes \theta^{-\frac{1}{2}}_{B'}U^{T\dagger}(m)_{B'E_B}\theta^{\frac{1}{2}}_{B'E_B})
\end{eqnarray*}
Let $D(m)_{BB'E_B}$ be the POVM element corresponding to message $m$. Then the probability of success, for a given message $m$, is 
\begin{eqnarray*}
\Tr\left(D(m)_{BB'E_B}(\id_A\otimes\theta^{\frac{1}{2}}_{B'E_B}U^T(m)_{B'E_B}\theta^{-\frac{1}{2}}_{B'})\cN_{A \to B}(\ketbra{\theta}_{AB'})\otimes \Phi_{E_B}(\id_A\otimes \theta^{-\frac{1}{2}}_{B'}U^{T\dagger}(m)_{B'E_B}\theta^{\frac{1}{2}}_{B'E_B})\right)
\end{eqnarray*}
}
We will prove the upper bound for uniform distribution over the messages. Fix a quantum state $\sigma_B$. Let $\psi_{MAE_B}$ be the quantum state after Alice's encoding. There exists a register $F$ that purifies $\psi_{MAE_B}$ into the pure state $\ket{\psi}_{MAE_BF}$. Let $\rho_{MBE_BF}$ be the quantum state after the action of the channel and $\phi_{MM'}= \cD(\rho_{MBE_B})$.  From Facts \ref{fact:dhallstates} and \ref{fact:monotonequantumoperation}, we have 
\begin{eqnarray*}
R &\leq& \dzeroseps{\eps}(\phi_{MM'}\|\phi_M \otimes \cD(\sigma_B \otimes \rho_{E_B}))\leq \dzeroseps{\eps}(\rho_{MBE_B}\|\rho_M \otimes \sigma_B \otimes \rho_{E_B}) \\ &=& \dzeroseps{\eps}(\rho_{MBE_B}\|\rho_{ME_B} \otimes \sigma_B) \leq \dzeroseps{\eps}(\rho_{MBE_BF}\|\rho_{ME_BF} \otimes \sigma_B),
\end{eqnarray*}
 where we have used the facts that $\rho_{ME_B} = \rho_M \otimes \rho_{E_B}$ and $\phi_M=\rho_M$.
Since register $B$ is obtained by an action of the channel $\cN_{A\to B}$, we have
$$R \leq \dzeroseps{\eps}(\cN_{A\to B}(\psi_{MAE_BF})\|\psi_{ME_BF} \otimes \sigma_B).$$ Setting $B' \defeq ME_BF$ and optimizing over all $\sigma_B$, we conclude the converse.
\end{proof}

Following achievability was shown in \cite{AnshuJW17CC}, which is near optimal.
\begin{theorem}[\cite{AnshuJW17CC}]
\label{theo:eaptopachieve}
Fix a quantum channel $\cN_{A\to B}$ and $\eps, \delta\in (0,1)$. There exists an $(R, 2\eps + \delta)$-entanglement assisted code for this channel if 
$$R \leq \max_{\psi_{AB'}}\dzeroseps{\eps + \delta}(\cN_{A\to B}(\psi_{AB'})\|\cN_{A\to B}(\psi_A)\otimes \psi_{B'}) - \log\frac{1}{\delta}.$$ 
\end{theorem}
The error of $2\eps + \delta$ can be improved to $\eps+\delta$, by tuning the parameter $c$ in Hayashi-Nagaoka inequality (Fact \ref{haynag}), as noted in \cite{QiWW17}.

\vspace{0.1in}

\noindent {\bf Success probability for entanglement assisted communication over noiseless channel.} Now, we recover the result in \cite[Theorem 1.1]{NayakS02} for one way protocols, as an application of Theorem \ref{entptopconverse}. 
\begin{corollary}
For any $(R, \eps)$- entanglement assisted code for the identity channel $\cN_{A\to A}(\rho_A) = \rho_A$, it holds that 
$$1-\eps \leq \frac{|A|^2}{2^R}.$$
\end{corollary}
\begin{proof}
We apply Theorem \ref{entptopconverse} with $\sigma_A = \frac{\id_A}{|A|}$ to obtain
$$R\leq \max_{\ket{\psi}_{AB'}}\dzeroseps{\eps}(\ketbra{\psi}_{AB'}\|\frac{\id_A}{|A|}\otimes \psi_{B'}).$$
Let $\ket{\psi}_{AB'} = \sum_i \lambda_i\ket{i}_A\ket{i}_{B'}$ be the Schmidt decomposition of $\ket{\psi}_{AB'}$ such that $\sum_i\lambda_i^2=1$ and let $\Pi' \defeq \sum_i \ketbra{i}_A\otimes \ketbra{i}_{B'}$. It holds that $\Pi'\ket{\psi}_{AB'} = \ket{\psi}_{AB'}$. From Fact \ref{fact:puredheps}, let $\ket{\Pi}$ (with some abuse of notation) be the rank one operator achieving the optimum for $\dzeroseps{\eps}(\ketbra{\psi}_{AB'}\|\frac{\id_A}{|A|}\otimes \psi_{B'})$. We recall that $\ket{\Pi}$ need not be normalized. Since $\bra{\Pi}\ket{\psi}_{AB'} = \bra{\Pi}\Pi'\ket{\psi}_{AB'}$ and 
$$\bra{\Pi}\Pi'\frac{\id_A}{|A|}\otimes \psi_{B'}\Pi'\ket{\Pi} = \bra{\Pi}\frac{\id_A}{|A|}\otimes \psi_{B'}\Pi'\ket{\Pi} \leq \bra{\Pi}\frac{\id_A}{|A|}\otimes \psi_{B'}\ket{\Pi},$$ we have that $\Pi'\ket{\Pi} = \ket{\Pi}$. Thus, we expand $\ket{\Pi} = \sum_i a_i \ket{i}_A\ket{i}_{B'}$ such that $\sum_i a_i^2 \leq 1$. The condition $|\bra{\Pi}\ket{\psi}_{AB'}|^2\geq 1-\eps$ translates to
$|\sum_i a_i \lambda_i|^2 \geq 1-\eps$. Further, 
$$\bra{\Pi}\frac{\id_A}{|A|}\otimes \psi_{B'}\ket{\Pi} = \frac{1}{|A|}\sum_i |a_i\lambda_i|^2.$$ By Cauchy-Schwartz inequality, 
$$1-\eps \leq |\sum_i a_i \lambda_i|^2 \leq |A|\sum_i |a_i\lambda_i|^2 \implies \frac{1-\eps}{|A|}\leq \sum_i |a_i\lambda_i|^2.$$ Hence, it holds that $$\bra{\Pi}\frac{\id_A}{|A|}\otimes \psi_{B'}\ket{\Pi} \geq \frac{1-\eps}{|A|^2},$$ for any feasible choice of $a_i, \lambda_i$. The inequality is achieved when $\lambda_i = \frac{1}{\sqrt{|A|}}$ and $a_i = \sqrt{\frac{1-\eps}{|A|}}$, which also satisfies the constraints $\sum_i \lambda_i^2=1, \sum_i a_i^2\leq 1, |\sum_i a_i\lambda_i|^2\geq 1-\eps$. Hence, we conclude that $$R\leq \max_{\ket{\psi}_{AB'}}\dzeroseps{\eps}(\ketbra{\psi}_{AB'}\|\frac{\id_A}{|A|}\otimes \psi_{B'}) \leq \log \frac{|A|^2}{1-\eps}.$$ 
\end{proof}

\subsection{Quantum channel with state}

Alice wants to communicate a classical message $M$ chosen from $[2^R]$ to Bob over a quantum channel $\cN_{AS \to B }$ such that Bob is able to decode the correct message with probability at least $1-\eps$. Alice shares entanglement $\ket{\tau}_{SS'}$ with the channel as well. This model in the classical setting is known as the {\em Gel'fand-Pinsker} channel.
\begin{definition}
Let $\ket{\theta}_{E_AE_B}$ be the shared entanglement between Alice and Bob and let $\ket{\tau}_{SS'}$ be the state shared between Alice and Channel. An $(R, \eps )$-entanglement assisted code for the quantum channel $\cN_{ AS \to B}$ consists of 
\begin{itemize}
\item An encoding operation $\cE: ME_A S' \rightarrow A$ for Alice. 
\item A decoding operation $\cD : B E_B\rightarrow M'$ for Bob, with $M'\equiv M$ being the output register such that for all $m \in [2^R]$,
\begin{equation*}
\Pr(M'\neq m| M=m)\leq \eps.
\end{equation*}
\end{itemize}
\end{definition}

We have the following converse.

\begin{theorem}
\label{gelpinentconv}
Fix a quantum channel $\cN_{AS \to B}$ with state $\tau_S$ and an $\eps\in (0,1)$. For every $(R, \eps)$-entanglement assisted code for this channel, it holds that
$$R \leq \max_{\psi_{ASB'}: \psi_{SB'} = \tau_S\otimes \psi_{B'}}\min_{\sigma_B}\dzeroseps{\eps}(\cN_{AS\to B}(\psi_{ASB'})\|\psi_{B'} \otimes \sigma_B).$$
\end{theorem}
\begin{proof}
We will prove the upper bound for uniform distribution over the messages. Fix a quantum state $\sigma_B$. Let $\psi_{MASE_B}$ be the quantum state after Alice's encoding and $\rho_{MBE_B}$ be the quantum state after the action of the channel. Let $\phi_{MM'} = \cD(\rho_{MBE_B})$. From Facts \ref{fact:dhallstates} and \ref{fact:monotonequantumoperation}, 
\begin{eqnarray*}
R &\leq& \dzeroseps{\eps}(\phi_{MM'}\|\phi_M \otimes \cD(\sigma_{B}\otimes \rho_{E_B})) \leq \dzeroseps{\eps}(\rho_{MBE_B}\|\rho_M \otimes \sigma_{B}\otimes \rho_{E_B})\\ &=& \dzeroseps{\eps}(\rho_{MBE_B}\|\rho_{ME_B} \otimes \sigma_{B}),
\end{eqnarray*}
 where we have used the facts that $\rho_{ME_B}= \rho_M \otimes \rho_{E_B}$ and $\rho_M=\phi_M$. Now, observe that $\cN_{AS\to B}(\psi_{MASE_B}) = \rho_{MBE_B}$ and $\psi_{MSE_B} = \psi_{ME_B}\otimes \tau_S$. Setting $B' = ME_B$, we conclude that 
$$R \leq \dzeroseps{\eps}(\cN_{AS\to B}(\psi_{ASB'})\|\psi_{B'} \otimes \sigma_B),$$ where $\psi_{SB'} = \tau_S\otimes \psi_{B'}$.
\end{proof}

As a corollary of above converse, we obtain the following converse statement, which matches (up to some constants) the achievability result given in \cite[Theorem 5]{AnshuJW17CC}. 
\begin{corollary}
Fix a quantum channel $\cN_{AS \to B}$ with state $\tau_S$ and an $\eps\in (0,1)$. For every $(R, \eps)$-entanglement assisted code for this channel, it holds that
$$R \leq \max_{\psi_{ASB'}: \psi_{S} = \tau_S}\left(\min_{\sigma_B}\dzeroseps{\eps}(\cN_{AS\to B}(\psi_{ASB'})\|\psi_{B'} \otimes \sigma_B) - \dmax{\psi_{SB'}}{\psi_S\otimes \psi_{B'}}\right).$$
\end{corollary}
\begin{proof}
If $\psi_{SB'}=\psi_S\otimes \psi_{B'}$, then $\dmax{\psi_{SB'}}{\psi_S\otimes \psi_{B'}}=0$. Thus, the optimization in above statement is over a larger set, as compared to that given in Theorem \ref{gelpinentconv}. 
\end{proof}
The utility of \cite[Theorem 5]{AnshuJW17CC} is that it yields a single letter expression in the asymptotic and i.i.d. setting, as shown in \cite{Frederic10} using different techniques. It is also possible to directly achieve the bound given in Theorem \ref{gelpinentconv}, as we show below. The utility of this bound is that it is of the form similar to that for multiple access channel and broadcast channel, both of which are one-shot optimal.
\begin{theorem}
\label{theo:qgelpinachieve}
Fix a quantum channel $\cN_{AS \to B}$ with state $\tau_S$ and $\eps, \delta\in (0,1)$. There exists an $(R, \eps+2\delta)$-entanglement assisted code for this channel, if  
$$R \leq \max_{\psi_{ASB'}: \psi_{SB'} = \tau_S\otimes \psi_{B'}}\dzeroseps{\eps}(\cN_{AS\to B}(\psi_{ASB'})\|\cN_{AS\to B}(\psi_{AS})\otimes \psi_{B'}) - \log\frac{4\eps}{\delta^2}.$$
\end{theorem}
\begin{proof}
Fix a quantum state $\psi_{ASB'}$ achieving the optimum above such that $\psi_{SB'}=\tau_S\otimes \psi_{B'}$. Let $\ket{\psi'}_{EB'}$ be a purification of $\psi_{B'}$. Alice and Bob share $2^R$ copies of $\ket{\psi'}_{EB'}$ in registers $E_1B'_1, \ldots E_{2^R}B'_{2^R}$. Let $\ket{\psi}_{AUSB'}$ be a purification of $\psi_{ASB'}$. Let $W: \cH_{S'E} \rightarrow \cH_{AU}$ be an isometry such that $W\ket{\psi'}_{EB'}\otimes \ket{\tau}_{S'S} = \ket{\psi}_{AUSB'}$.

\vspace{0.1in}

\noindent {\bf Encoding:} To send the message $m \in [2^R]$, Alice prepares the pure state $\ket{\psi}_{AUSB'_m}$ by applying the isometry $W$ on the registers $S'E_m$ and sends register $A$ through the channel. 

\vspace{0.1in}

\noindent {\bf Decoding and error analysis:} Bob performs the position-based decoding strategy across the $B'$ registers. Let $\Pi_{BB'}$ be the operator achieving the optimum in the definition of 
$\dzeroseps{\eps}(\cN_{AS\to B}(\psi_{ASB'})\|\cN_{AS\to B}(\psi_{AS})\otimes \psi_{B'})$.  Define 
$$\Lambda(m)\defeq \mathrm{I}_{B'_1} \otimes \mathrm{I}_{B'_2} \otimes \cdots \Pi_{BB'_m} \otimes \cdots \otimes \mathrm{I}_{B'_{2^{R}}},$$ 
and 
$$\Omega(m) \defeq \left(\sum_{ m' \in [1:2^{R}]} \Lambda(m')\right)^{-\frac{1}{2}}\Lambda(m)\left(\sum_{m' \in [1:2^{R_1}]} \Lambda(m')\right)^{-\frac{1}{2}}.$$
Bob applies the measurement $\{\Omega(1), \ldots \Omega(2^{R}), \id - \sum_{m}\Omega(m)\}$ to decode $m$.

\vspace{0.1in}

\noindent {\bf Error analysis:} Employing Hayashi-Nagaoka inequality (Fact \ref{haynag}), we have 
\begin{eqnarray*}
&&\Pr\{M'\neq m\mid M= m\} \\ && \leq (1+c) \Tr((\id-\Pi_{BB'})\cN_{AS\to B}(\psi_{ASB'})) + (2+c+\frac{1}{c})\cdot 2^{R_1}\Tr(\Pi_{BB'}\cN_{AS\to B}(\psi_{AS})\otimes \psi_{B'})\\
&& \leq (1+c)\eps + \frac{4}{c}2^{R_1-\dzeroseps{\eps}(\cN_{AS\to B}(\psi_{ASB'})\|\cN_{AS\to B}(\psi_{AS})\otimes \psi_{B'})}\\
&& \leq \eps+2\delta,
\end{eqnarray*}
where we choose $c = \frac{\delta}{\eps}$. 

This completes the proof.
\end{proof}

\subsection{Broadcast quantum channel}

Alice wishes to communicate message pair $(m_1,m_2)$ simultaneously to Bob and Charlie over a quantum broadcast channel, where $m_1$ is intended for Bob and $m_2$ is intended for Charlie, such that both Bob and Charlie output the correct message with probability at least $1-\eps$. 
\begin{definition}
Let $\ket{\theta}_{E_{A_1}E_B}$ and $\ket{\theta}_{E_{A_2}E_C}$ be the shared entanglement between Alice and Bob and Alice and Charlie respectively. An $(R_1, R_2, \eps_1, \eps_2)$ entanglement assisted code for the quantum broadcast channel $\cN_{ A \to BC}$ consists of 
\begin{itemize}
\item An encoding operation $\cE: M_1M_2E_{A_1}E_{A_2} \rightarrow A$ for Alice.  
\item A pair of decoding operations $(\cD_B,\cD_C)$,  $\cD_B : B E_B\rightarrow M'_1$ and $\cD_C : C E_C\rightarrow M'_2$, with $(M_1',M_2')\equiv (M_1,M_2)$ being the output registers, such that for all $(m_1,m_2) \in [2^{R_1}]\times [2^{R_2}]$
\begin{equation*}
\Pr(M'_1\neq m_1| M_1 = m_1)\leq \eps_1, \Pr(M'_2\neq m_2| M_2 = m_2)\leq \eps_2.
\end{equation*}
\end{itemize}
\end{definition}

We have the following converse.

\begin{theorem}
\label{theo:broadcastconv}
Fix a quantum channel $\cN_{A\to BC}$ and $\eps_1, \eps_2 \in (0,1)$. For any $(R_1, R_2, \eps_1, \eps_2)$- entanglement assisted code for this channel, there exist registers $B',C'$ and a quantum state $\psi_{AB'C'}$ satisfying $\psi_{B'C'}= \psi_{B'}\otimes \psi_{C'}$ such that
$$R_1 \leq \min_{\sigma_B}\dzeroseps{\eps_1}(\Tr_C\cN_{A\to BC}(\psi_{AB'})\|\sigma_B \otimes \psi_{B'}),$$
$$R_2 \leq \min_{\tau_C}\dzeroseps{\eps_2}(\Tr_B\cN_{A\to BC}(\psi_{AC'})\|\tau_C \otimes \psi_{C'}).$$ 
\end{theorem}
\begin{proof}
We will  prove the upper bound for uniform distribution over the messages. Fix quantum states $\sigma_B, \tau_C$. Let $\psi_{M_1M_2AE_BE_C}$ be the quantum state after Alice's encoding and $\rho_{M_1M_2BCE_BE_C}$ be the quantum state after the action of the channel. Let $\phi^B_{M_1M'_1}\defeq \cD_B(\rho_{M_1BE_B})$ and $\phi^C_{M_2M'_2}\defeq \cD_C(\rho_{M_2CE_C})$. From Facts \ref{fact:dhallstates} and \ref{fact:monotonequantumoperation},
$$R_1 \leq \dzeroseps{\eps_1}(\phi^B_{M_1M'_1}\|\phi^B_{M_1} \otimes \cD(\sigma_{B}\otimes \rho_{E_B})) \leq \dzeroseps{\eps_1}(\rho_{M_1BE_B}\|\rho_{M_1} \otimes \sigma_{B}\otimes \rho_{E_B})= \dzeroseps{\eps_1}(\rho_{M_1BE_B}\|\rho_{M_1E_B} \otimes \sigma_{B}),$$ where we have used the facts that $\rho_{M_1E_B}= \rho_{M_1} \otimes \rho_{E_B}$ and $\phi^B_{M_1}=\rho_{M_1}$. Similarly, 
$$R_2\leq \dzeroseps{\eps_2}(\rho_{M_2CE_C}\|\rho_{M_2E_C} \otimes \tau_{C}).$$ Observe that $\cN_{A\to BC}(\psi_{M_1M_2AE_BE_C}) = \rho_{M_1M_2BCE_BE_C}$ and $\psi_{M_1M_2E_BE_C} = \psi_{M_1E_B}\otimes \psi_{M_2E_C}$. Define $B'\defeq M_1E_B$ and $C'\defeq M_2E_C$. Thus, we conclude that 
$$R_1 \leq \dzeroseps{\eps_1}(\Tr_C\cN_{A\to BC}(\psi_{AB'})\|\sigma_B \otimes \psi_{B'}), \quad R_2 \leq \dzeroseps{\eps_2}(\Tr_B\cN_{A\to BC}(\psi_{AC'})\|\tau_C \otimes \psi_{C'}),$$ where $\psi_{B'C'}= \psi_{B'}\otimes \psi_{C'}$.
\end{proof}

As a corollary, we obtain the following converse result, which shows the one-shot near optimality of the bound for quantum broadcast channel given in \cite{AnshuJW17CC}. The fact that the following corollary follows from Theorem \ref{theo:broadcastconv} is implicit in the asymptotic and i.i.d. converse of \cite[Theorem 3]{DupuisHL10}.

\begin{corollary}
Fix a quantum channel $\cN_{A\to BC}$ and $\eps_1, \eps_2 \in (0,1)$. For any $(R_1, R_2, \eps_1, \eps_2)$- entanglement assisted code for this channel, there exist registers $B',C'$ and a quantum state $\psi_{AB'C'}$ such that
\begin{eqnarray*}
R_1 &\leq& \min_{\sigma_B}\dzeroseps{\eps_1}(\Tr_C\cN_{A\to BC}(\psi_{AB'})\|\sigma_B \otimes \psi_{B'}),\\
R_2 &\leq& \min_{\tau_C}\dzeroseps{\eps_2}(\Tr_B\cN_{A\to BC}(\psi_{AC'})\|\tau_C \otimes \psi_{C'}), \\
R_1+R_2 &\leq& \min_{\sigma_B}\dzeroseps{\eps_1}(\Tr_C\cN_{A\to BC}(\psi_{AB'})\|\sigma_B \otimes \psi_{B'}) +\\ &&\min_{\tau_C}\dzeroseps{\eps_2}(\Tr_B\cN_{A\to BC}(\psi_{AC'})\|\tau_C \otimes \psi_{C'}) - \dmax{\psi_{B'C'}}{\psi_{B'}\otimes \psi_{C'}}.
\end{eqnarray*}
\end{corollary}
\begin{proof}
The proof follows by relaxing the constraint $\psi_{B'C}=\psi_{B'}\otimes \psi_{C'}$ in Theorem \ref{theo:broadcastconv}.
\end{proof}

We have the following achievability result.

\begin{theorem}
\label{theo:qbroadachieve}
Fix a quantum channel $\cN_{A\to BC}$ and $\eps_1, \eps_2, \delta \in (0,1)$. For every quantum state $\psi_{AB'C'}$ satisfying $\psi_{B'C'} = \psi_{B'}\otimes \psi_{C'}$, there exists an $(R_1, R_2, \eps_1+\delta, \eps_2 + \delta)$- entanglement assisted code for this channel, if 
$$R_1 \leq \dzeroseps{\eps_1}(\Tr_C\cN_{A\to BC}(\psi_{AB'})\|\Tr_C\cN_{A\to BC}(\psi_A) \otimes \psi_{B'}) - \log\frac{4\eps_1}{\delta^2},$$
$$R_2 \leq \dzeroseps{\eps_2}(\Tr_B\cN_{A\to BC}(\psi_{AC'})\|\Tr_B\cN_{A\to BC}(\psi_A) \otimes \psi_{C'}) - \log\frac{4\eps_2}{\delta^2}.$$ 
\end{theorem} 
\begin{proof}
Let $\ket{\psi}_{UAB'C'}$ be a purification of $\psi_{AB'C'}$. Let $\ket{\kappa}_{EB'}$ be a purification of $\psi_{B'}$ and $\ket{\kappa}_{FC'}$ be a purification of $\psi_{C'}$. Let $W: \cH_{EF} \to \cH_{UA}$ be an isometry such that $W\ket{\kappa}_{EB'}\otimes \ket{\kappa}_{FC'} = \ket{\psi}_{UAB'C'}$. Alice and Bob share $2^{R_1}$ copies of the quantum state $\ket{\kappa}_{EB'}$ in registers $E_1B'_1, \ldots E_{2^{R_1}}B'_{2^{R_1}}$. Alice and Charlie share $2^{R_2}$ copies of the quantum state $\ket{\kappa}_{FC'}$ in registers $F_1C'_1, \ldots F_{2^{R_1}}C'_{2^{R_1}}$. 

\vspace{0.1in}

\noindent {\bf Encoding:} To send the message pair $(m_1, m_2)$, Alice prepares the pure state $\ket{\psi}_{UAB'_{m_1}C'_{m_2}}$ by applying the isometry $W$ on the registers $E'_{m_1}, F'_{m_2}$ and sends the register $A$ through the channel. 

\vspace{0.1in}

\noindent {\bf Decoding:} Bob and Charlie apply the position based decoding strategy. Let $\Pi_{BB'}$ be the operator that achieves the optimum in $\dzeroseps{\eps_1}(\Tr_C\cN_{A\to BC}(\psi_{AB'})\|\Tr_C\cN_{A\to BC}(\psi_A) \otimes \psi_{B'})$ and $\Pi_{CC'}$ be the operator that achieves the optimum in $\dzeroseps{\eps_2}(\Tr_B\cN_{A\to BC}(\psi_{AC'})\|\Tr_B\cN_{A\to BC}(\psi_A) \otimes \psi_{C'})$. 
Define 
$$\Lambda_B(m_1)\defeq \mathrm{I}_{B'_1} \otimes \mathrm{I}_{B'_2} \otimes \cdots \Pi_{BB'_{m_1}} \otimes \cdots \otimes \mathrm{I}_{B'_{2^{R_1}}},$$ 
$$\Lambda_C(m_2)\defeq \mathrm{I}_{C'_1} \otimes \mathrm{I}_{C'_2} \otimes \cdots \Pi_{CC'_{m_2}} \otimes \cdots \otimes \mathrm{I}_{C'_{2^{R_2}}},$$
and 
$$\Omega_B(m_1) \defeq \left(\sum_{ m'_1 \in [1:2^{R_1}]} \Lambda_B(m'_1)\right)^{-\frac{1}{2}}\Lambda_B(m_1)\left(\sum_{m'_1 \in [1:2^{R_1}]} \Lambda_B(m'_1)\right)^{-\frac{1}{2}},$$
$$\Omega_C(m_2) \defeq \left(\sum_{ m'_2 \in [1:2^{R_2}]} \Lambda_C(m'_2)\right)^{-\frac{1}{2}}\Lambda_C(m_2)\left(\sum_{m'_2 \in [1:2^{R_2}]} \Lambda_C(m'_2)\right)^{-\frac{1}{2}}.$$
Bob applies the measurement $\{\Omega_B(1), \ldots \Omega_B(2^{R_1}), \id - \sum_{m_1}\Omega_B(m_1)\}$ to decode $m_1$. Charlie applies the measurement $\{\Omega_C(1), \ldots \Omega_C(2^{R_2}), \id - \sum_{m_2}\Omega_C(m_2)\}$ to decode $m_2$.

\vspace{0.1in}

\noindent {\bf Error analysis:} Employing Hayashi-Nagaoka inequality (Fact \ref{haynag}), we have 
\begin{eqnarray*}
&&\Pr\{M'_1\neq m_1\mid M_1= m_1\} \\ && \leq (1+c) \Tr((\id-\Pi_{BB'})\Tr_C\cN_{A\to BC}(\psi_{AB'})) + (2+c+\frac{1}{c})\cdot 2^{R_1}\Tr(\Pi_{BB'}\Tr_C\cN_{A\to BC}(\psi_{A})\otimes \psi_{B'})\\
&& \leq (1+c)\eps_1 + \frac{4}{c}2^{R_1-\dzeroseps{\eps_1}(\Tr_C\cN_{A\to BC}(\psi_{AB'})\|\Tr_C\cN_{A\to BC}(\psi_A) \otimes \psi_{B'})}\\
&& \leq \eps_1+2\delta,
\end{eqnarray*}
where we choose $c = \frac{\delta}{\eps_1}$. Similarly, we have 
$$\Pr\{M'_2\neq m_2\mid M_2= m_2\} \leq \eps_2 + 2\delta.$$ This completes the proof.
\end{proof}

\subsection{Multiple access channel}

Alice wants to communicate a classical message $m_1$ chosen from $[2^{R_1}]$ to Charlie and Bob wants to communicate a classical message $m_2$ chosen from $[2^{R_2}]$ to Charlie, over a channel $\cN_{AB \to C}$. Alice shares entanglement with Charlie and Bob shares entanglement with Charlie. Alice and Bob do not share entanglement. This is the multiple access channel. 
\begin{definition}
Let $\ket{\theta}_{E_{A}E_{C_1}}$ and $\ket{\theta}_{E_BE_{C_2}}$ be the shared entanglement between Alice and Charlie and Bob and Charlie, respectively. An $(R_1, R_2, \eps_1, \eps_2)$- entanglement assisted code for the quantum multiple access channel $\cN_{AB \to C}$ consists of 
\begin{itemize}
\item Encoding operations $\cE_1: M_1E_{A} \rightarrow A$ and $\cE_2: M_2E_B\rightarrow B$.
\item A decoding operation $\cD: E_{C_1}E_{C_2}C \rightarrow M'_1M'_2$, with $M'_1\equiv M_1$ and $M'_2\equiv M_2$ such that for all $(m_1, m_2) \in [2^{R_1}]\times [2^{R_2}]$,
$$\Pr(M'_1\neq m_1| M_1 = m_1)\leq \eps_1, \Pr(M'_2\neq m_2| M_2 = m_2)\leq \eps_2.$$ 
\end{itemize}  
\end{definition}
We note that the definition of error above is closely related to the definition of error as  $\Pr(M'_1, M'_2\neq m_1, m_2| M_1, M_2 = m_1, m_2)\leq \eps$ through
\begin{eqnarray*}
\max_i\{\Pr(M'_i\neq m_i| M_i = m_i)\}&\leq& \Pr(M'_1, M'_2\neq m_1, m_2| M_1, M_2 = m_1, m_2) \\ &\leq& \Pr(M'_1\neq m_1| M_1 = m_1) + \Pr(M'_2\neq m_2| M_2 = m_2).
\end{eqnarray*}

We have the following converse.
\begin{theorem}
\label{multaccentconv}
Fix a quantum channel $\cN_{AB\to C}$ and $\eps_1, \eps_2\in (0,1)$. For every $(R_1, R_2, \eps_1, \eps_2)$-entanglement assisted code for this channel, there exist registers $A',A'', B', B''$ and a quantum state $\psi_{ABA'B'A''B''}$ satisfying $\psi_{ABA'B'A''B''} = \psi_{AA'A''}\otimes \psi_{BB'B''}$ and $\psi_{A'B'A''B''} = \psi_{A'}\otimes \psi_{A''}\otimes \psi_{B'}\otimes \psi_{B''}$ such that 
$$R_1 \leq \dzeroseps{\eps_1}(\rho_{A'CA''B''}\| \rho_{CA''B''}\otimes \rho_{A'}),$$
and
$$R_2 \leq \dzeroseps{\eps_2}(\rho_{B'CA''B''}\| \rho_{CA''B''}\otimes \rho_{B'}),$$
where $\rho_{CA'B'} \defeq \cN_{AB\to C}(\psi_{ABA'B'})$.
\end{theorem}
\begin{proof}
We will prove the upper bound for uniform distribution over the messages. Let $\psi_{M_1M_2ABE_{C_1}E_{C_2}}$ be the quantum state after the operations of Alice and Bob. It holds that $\psi_{M_1M_2ABE_{C_1}E_{C_2}} = \psi_{M_1AE_{C_1}}\otimes \psi_{M_2BE_{C_2}}$. Let $\rho_{M_1M_2CE_{C_1}E_{C_2}}$ be the quantum state after the action of the channel. Let $\phi_{M_1M_2M'_1M'_2} = \cD(\rho_{M_1M_2CE_{C_1}E_{C_2}})$. From Facts \ref{fact:dhallstates} and \ref{fact:monotonequantumoperation},
\begin{eqnarray*}
R_1 &\leq& \dzeroseps{\eps_1}(\phi_{M_1M'_1M'_2}\|\phi_{M_1}\otimes \cD(\rho_{CE_{C_1}E_{C_2}})) \leq \dzeroseps{\eps_1}(\rho_{M_1CE_{C_1}E_{C_2}}\| \rho_{M_1}\otimes \rho_{CE_{C_1}E_{C_2}}).
\end{eqnarray*}
Similarly,
$$R_2 \leq \dzeroseps{\eps_2}(\rho_{M_2E_{C_2}CE_{C_1}}\| \rho_{M_2}\otimes \rho_{CE_{C_2}E_{C_1}}).$$
\suppress{Finally,
\begin{eqnarray*}
\log |M_1||M_2| &=& R_1+R_2 \leq \dzeroseps{\eps_1+\eps_2}(\phi_{M_1M_2M'_1M'_2}\|\phi_{M_1M_2}\otimes \cD(\rho_{C} \otimes \rho_{E_{C_1}E_{C_2}}))\\ 
&\leq& \dzeroseps{\eps_1+\eps_2}(\rho_{M_1M_2E_{C_1}E_{C_2}C}\| \rho_{M_1M_2E_{C_1}E_{C_2}}\otimes \rho_C)\\
&\leq& \frac{\relent{\rho_{M_1M_2E_{C_1}E_{C_2}C}}{\rho_{M_1M_2E_{C_1}E_{C_2}}\otimes \rho_C}}{1-\eps_1-\eps_2}\\
&\leq& \frac{\log|C|}{1-\eps_1-\eps_2},
\end{eqnarray*}
where we have used $$\rho_{M_1M_2E_{C_1}E_{C_2}} = \rho_{M_1}\otimes \rho_{M_2}\otimes \rho_{E_{C_1}}\otimes \rho_{E_{C_2}}.$$ Thus, $|M_1||M_2|\leq |C|^{\frac{1}{1-\eps_1-\eps_2}}$. }We observe the relations $\cN_{AB\to C}(\psi_{M_1M_2ABE_{C_1}E_{C_2}}) = \rho_{M_1M_2E_{C_2}CM_1E_{C_1}}$,  and define $A' \defeq M_1, A'' \defeq E_{C_1}$, $B'\defeq M_2, B''\defeq E_{C_2}$. This concludes the proof.
\end{proof}

It is also possible to obtain a one-shot version of the converse given in \cite{HsiehDW08}, as follows. 
\begin{theorem}
Fix a quantum channel $\cN_{AB\to C}$ and $\eps_1, \eps_2\in (0,1)$. For every $(R_1, R_2, \eps_1, \eps_2)$-entanglement assisted code for this channel, there exist registers $A', B'$ and a pure quantum state $\ket{\psi}_{ABA'B'}$ satisfying $\ket{\psi}_{ABA'B'} = \ket{\psi}_{AA'}\otimes \ket{\psi}_{BB'}$, such that 
$$R_1 \leq \dzeroseps{\eps_1}(\rho_{CA'B'}\| \rho_{CB'}\otimes \rho_{A'}),$$
$$R_2 \leq \dzeroseps{\eps_2}(\rho_{CA'B'}\| \rho_{CA'}\otimes \rho_{B'})$$
and
$$R_1+R_2 \leq \dzeroseps{\eps_2}(\rho_{CA'B'}\| \rho_{C} \otimes \rho_{A'}\otimes \rho_{B'}),$$
where $\rho_{CA'B'} \defeq \cN_{AB\to C}(\psi_{ABA'B'})$.
\end{theorem}
\begin{proof}
We will prove the upper bound for uniform distribution over the messages. Let $\psi_{M_1M_2ABE_{C_1}E_{C_2}}$ be the quantum state after the operations of Alice and Bob. It holds that $\psi_{M_1M_2ABE_{C_1}E_{C_2}} = \psi_{M_1AE_{C_1}}\otimes \psi_{M_2BE_{C_2}}$. Let $F_1, F_2$ be registers such that $\ket{\psi}_{M_1AE_{C_1}F_1}$ purifies $\psi_{M_1AE_{C_1}}$ and $\ket{\psi}_{M_2BE_{C_2}F_2}$ purifies $\psi_{M_2BE_{C_2}}$. Let $\rho_{M_1M_2CE_{C_1}F_1E_{C_2}F_2}$ be the quantum state after the action of the channel. Let $\phi_{M_1M_2M'_1M'_2} = \cD(\rho_{M_1M_2CE_{C_1}E_{C_2}})$. From Facts \ref{fact:dhallstates} and \ref{fact:monotonequantumoperation},
\begin{eqnarray*}
R_1 &\leq& \dzeroseps{\eps_1}(\phi_{M_1M'_1M'_2}\|\phi_{M_1}\otimes \cD(\rho_{E_{C_1}} \otimes \rho_{CE_{C_2}})) \leq \dzeroseps{\eps_1}(\rho_{M_1CE_{C_1}E_{C_2}}\| \rho_{M_1}\otimes \rho_{E_{C_1}}\otimes \rho_{CE_{C_2}})\\ &\leq& \dzeroseps{\eps_1}(\rho_{M_1CE_{C_1}F_1E_{C_2}F_2}\| \rho_{M_1E_{C_1}F_1}\otimes \rho_{CE_{C_2}F_2}) \leq \dzeroseps{\eps_1}(\rho_{M_1CE_{C_1}F_1M_2E_{C_2}F_2}\| \rho_{M_1E_{C_1}F_1}\otimes \rho_{CM_2E_{C_2}F_2}),
\end{eqnarray*}
where we have used $\rho_{M_1E_{C_1}} = \rho_{M_1}\otimes \rho_{E_{C_1}}$. Similarly,
$$R_2 \leq \dzeroseps{\eps_2}(\rho_{M_2E_{C_2}F_2CM_1E_{C_1}F_1}\| \rho_{M_2E_{C_2}F_2}\otimes \rho_{CM_1E_{C_1}F_1}).$$
Further using Facts \ref{fact:dhallstates} and \ref{fact:monotonequantumoperation},
\begin{eqnarray*}
R_1+R_2 &\leq& \dzeroseps{\eps_1+\eps_2}(\phi_{M_1M_2M'_1M'_2}\|\phi_{M_1}\otimes \phi_{M_2}\otimes \cD(\rho_{C} \otimes \rho_{E_{C_1}} \otimes \rho_{E_{C_2}}))\\ &\leq& \dzeroseps{\eps_1}(\rho_{M_1CE_{C_1}E_{C_2}}\| \rho_{M_1}\otimes \rho_{M_2}\otimes \rho_{C} \otimes \rho_{E_{C_1}} \otimes \rho_{E_{C_2}})\\ &=& \dzeroseps{\eps_1}(\rho_{M_1CE_{C_1}E_{C_2}}\| \rho_{M_1E_{C_1}}\otimes \rho_{M_2E_{C_2}}\otimes \rho_{C})\\ &\leq& \dzeroseps{\eps_1}(\rho_{M_1CE_{C_1}F_1E_{C_2}F_2}\| \rho_{M_1E_{C_1}F_1}\otimes \rho_{M_2E_{C_2}F_2}\otimes \rho_{C}).
\end{eqnarray*}
We observe the relations $\cN_{AB\to C}(\psi_{M_1M_2ABE_{C_1}F_1E_{C_2}F_2}) = \rho_{M_1M_2CE_{C_1}F_1E_{C_2}F_2}$,  and define $A' \defeq M_1E_{C_1}F_1$, $B'\defeq M_2E_{C_2}F_2$. This concludes the proof.
\end{proof}

While above converse has the utility that it involves optimization over registers of bounded dimensions, in contrast to converse in Theorem \ref{multaccentconv}, it is not clear how to achieve it without an appropriate notion of simultaneous decoding. On the other hand, we have the following achievability result, which is near-optimal with respect to the converse given in Theorem \ref{multaccentconv}, for either one of the error parameters. Furthermore, it does not require a simultaneous decoding strategy.

\begin{theorem}
\label{qmultaccessachieve}
Fix a quantum channel $\cN_{AB\to C}$ and $\eps_1, \eps_2, \delta \in (0,1)$. Let there be a quantum state $\psi_{ABA'B'A''B''}$ satisfying $\psi_{ABA'B'A''B''} = \psi_{AA'A''}\otimes \psi_{BB'B''}$ and $\psi_{A'B'A''B''} = \psi_{A'}\otimes \psi_{A''}\otimes \psi_{B'}\otimes \psi_{B''}$. There exists an $(R_1, R_2, \eps_1+2\delta, \eps_2 + 2\delta+3\sqrt{\eps_1+2\delta})$- entanglement assisted code and an $(R_1, R_2 , \eps_1 + 2\delta+3\sqrt{\eps_2+2\delta},\eps_2+2\delta)$- entanglement assisted code for this channel,  
$$R_1 \leq \dzeroseps{\eps_1}(\rho_{A'CA''B''}\| \rho_{CA''B''}\otimes \rho_{A'}) - \log\frac{4\eps_1}{\delta^2},$$
and
$$R_2 \leq \dzeroseps{\eps_2}(\rho_{B'CA''B''}\| \rho_{CA''B''}\otimes \rho_{B'}) - \log\frac{4\eps_2}{\delta^2},$$
where $\rho_{CA'B'A''B''} \defeq \cN_{AB\to C}(\psi_{ABA'B'A''B''})$.
\end{theorem}
\begin{proof}
Introduce registers $G_1, G_2$ such that $\ket{\psi}_{AA'A''G_1}$ and $\ket{\psi}_{BB'B''G_2}$ purify $\psi_{AA'A''}$ and $\psi_{BB'B''}$. Let $\rho_{CA'B'A''B''}\defeq \cN_{AB \to C}(\psi_{AA'A''}\otimes \psi_{BB'B''})$. Let $\ket{\kappa}_{E'A'}, \ket{\kappa}_{E''A''}, \ket{\kappa}_{F'B'}, \ket{\kappa}_{F''B''}$ be the purifications of $\psi_{A'}, \psi_{A''}, \psi_{B'}, \psi_{B''}$. Let $V_A: \cH_{E''E'}\rightarrow \cH_{AG_1}$ be an isometry such that $V_A\ket{\kappa}_{E'A'}\otimes \ket{\kappa}_{E''A''} = \ket{\psi}_{AA'A''G_1}$ and $V_B: \cH_{F''F'}\rightarrow \cH_{BG_2}$ be an isometry such that $V_B\ket{\kappa}_{F'B'}\otimes \ket{\kappa}_{F''B''} = \ket{\psi}_{BB'B''G_2}$.

Alice and Charlie share one copy of $\ket{\kappa}_{E''A''}$, where Alice holds $E''$ and Charlie holds $A''$, and $2^{R_1}$ copies of $\ket{\kappa}_{E'A'}$ in registers $E'_1, E'_2, \ldots E'_{2^{R_1}}$, where Alice holds $E'_1, \ldots E'_{2^{R_1}}$ and Charlie holds $A'_1, \ldots A'_{2^{R_1}}$. Bob and Charlie share one copy of $\ket{\kappa}_{F''B''}$, where Bob holds $F''$ and Charlie holds $B''$, and $2^{R_2}$ copies of $\ket{\kappa}_{F'B'}$ in registers $F'_1B'_1, \ldots F'_{2^{R_2}}B'_{2^{R_2}}$, where Bob holds $F'_1, \ldots F'_{2^{R_2}}$ and Charlie holds $A'_1, \ldots A'_{2^{R_1}}$.

\vspace{0.1in}

\noindent {\bf Encoding:} To send the message $m_1 \in [2^{R_1}]$, Alice applies the isometry $V_A$ on registers $E''E'_{m_1}$ to prepare the purification $\ket{\psi}_{AA''A'_{m_1}G_1}$ of $\psi_{A''}\otimes \psi_{A'_{m_1}}$. She sends $A$ through the channel. 

To send the message $m_2$, Bob applies an isometry $V_B$ on the registers $F''F'_{m_2}$ to prepare the purification $\ket{\psi}_{BB''B'_{m_2}G_2}$ of $\psi_{B''}\otimes \psi_{B'_{m_2}}$. He sends $B$ through the channel. 

\vspace{0.1in}

\noindent {\bf Decoding:} Charlie applies the position based decoding strategy \cite{AnshuJW17CC} independently across $A''$ registers and then $B''$ registers as follows. Let $\Pi_{CA''B''A'}$ be the operator achieving the optimum in $\dzeroseps{\eps_1}(\rho_{A'CA''B''}\| \rho_{CA''B''}\otimes \rho_{A'})$ and $\Pi_{CA''B''B'}$ be the operator achieving the optimum in $\dzeroseps{\eps_2}(\rho_{B'CA''B''}\| \rho_{CA''B''}\otimes \rho_{B'})$. Define 
$$\Lambda_A(m_1)\defeq \mathrm{I}_{A'_1} \otimes \mathrm{I}_{A'_2} \otimes \cdots \Pi_{CA''B''A'_m} \otimes \cdots \otimes \mathrm{I}_{A'_{2^{R_1}}},$$ 
$$\Lambda_B(m_1)\defeq \mathrm{I}_{B'_1} \otimes \mathrm{I}_{B'_2} \otimes \cdots \Pi_{CA''B''B'_m} \otimes \cdots \otimes \mathrm{I}_{B'_{2^{R_2}}},$$
and 
$$\Omega_A(m_1) \defeq \left(\sum_{ m'_1 \in [1:2^{R_1}]} \Lambda_A(m'_1)\right)^{-\frac{1}{2}}\Lambda_A(m_1)\left(\sum_{m'_1 \in [1:2^{R_1}]} \Lambda_A(m'_1)\right)^{-\frac{1}{2}},$$
$$\Omega_B(m_2) \defeq \left(\sum_{ m'_2 \in [1:2^{R_2}]} \Lambda_B(m'_2)\right)^{-\frac{1}{2}}\Lambda_B(m_2)\left(\sum_{m'_2 \in [1:2^{R_2}]} \Lambda_B(m'_2)\right)^{-\frac{1}{2}}.$$
Charlie applies first the measurement $\{\Omega_A(1), \ldots \Omega_A(2^{R_1}), \id - \sum_{m_1}\Omega_A(m_1)\}$ to decode $m_1$. Then he applies the measurement $\{\Omega_B(1), \ldots \Omega_B(2^{R_2}), \id - \sum_{m_2}\Omega_B(m_2)\}$ to decode $m_2$.

\vspace{0.1in}

\noindent {\bf Error analysis:} Following the argument in \cite{AnshuJW17CC} and employing Hayashi-Nagaoka inequality (Fact \ref{haynag}), we have 
\begin{eqnarray*}
&&\Pr\{M'_1\neq m_1\mid M_1= m_1\} \\ && \leq (1+c) \Tr((\id-\Pi_{CA''B''A'})\rho_{CA''B''A'}) + (2+c+\frac{1}{c})\cdot 2^{R_1}\Tr(\Pi_{CA''B''A'}\psi_{CA''B''}\otimes \psi_{A'})\\
&& \leq (1+c)\eps_1 + \frac{4}{c}2^{R_1-\dzeroseps{\eps_1}(\rho_{A'CA''B''}\| \rho_{CA''B''}\otimes \rho_{A'})}\\
&& \leq \eps_1+2\delta,
\end{eqnarray*}
where we choose $c = \frac{\delta}{\eps_1}$. 

Let $M''_2$ be the output if Charlie first performed the measurement $\{\Omega_B(1), \ldots \Omega_B(2^{R_2}), \id - \sum_{m_2}\Omega_B(m_2)\}$. Then we would have $\Pr\{M''_2 \neq m_2\mid M_2 = m_2\} \leq \eps_2+2\delta$.
From Fact \ref{gentlepovm}, we conclude that the purified distance between the global quantum states before and after Charlie's first measurement is at most $\sqrt{2\eps_1+4\delta}$. From Fact \ref{closestatesmeasurement}, we thus conclude that
$$\Pr\{M'_2 \neq m_2\mid M_2=m_2\} \leq (\sqrt{\Pr\{M''_2 \neq m_2\mid M_2=m_2\}} + \sqrt{2\eps_1+4\delta})^2 \leq \eps_2 + 2\delta + 3\sqrt{\eps_1+2\delta}.$$
By decoding $m_2$ before $m_1$, an alternate protocol can be obtained. This completes the proof.
\end{proof}

Above theorem has the limitation that the overall error scales as $O(\eps_1+\sqrt{\eps_2})$ or $O(\eps_2+\sqrt{\eps_1})$. We improve it in the following theorem., using the sequential decoding technique (Fact \ref{noncommutativebound}).

\begin{theorem}
\label{qmultaccessgood}
Fix a quantum channel $\cN_{AB\to C}$ and $\eps_1, \eps_2, \delta \in (0,1)$. Let there be a quantum state $\ket{\psi}_{ABA'B'A''B''}$ satisfying $\psi_{ABA'B'A''B''} = \psi_{AA'A''}\otimes \psi_{BB'B''}$ and $\psi_{A'B'A''B''} = \psi_{A'}\otimes \psi_{A''}\otimes \psi_{B'}\otimes \psi_{B''}$. There exists an $(R_1, R_2, 4(\eps_1+\eps_2 + 2\delta), 4(\eps_1+\eps_2 + 2\delta))$- entanglement assisted code for this channel, if
$$R_1 \leq \dzeroseps{\eps_1}(\rho_{A'CA''B''}\| \rho_{CA''B''}\otimes \rho_{A'}) - \log\frac{1}{\delta},$$
and
$$R_2 \leq \dzeroseps{\eps_2}(\rho_{B'CA''B''}\| \rho_{CA''B''}\otimes \rho_{B'}) - \log\frac{1}{\delta},$$
where $\rho_{CA'B'A''B''} \defeq \cN_{AB\to C}(\psi_{ABA'B'A''B''})$. In fact, the following upper bound holds for all $(m_1, m_2)$ in $[2^{R_1}]\times [2^{R_2}]$,
$$\Pr(M'_1, M'_2 = m_1, m_2\mid M_1, M_2 = m_1, m_2) \leq 4\cdot (\eps_1+\eps_2 + 2\delta).$$
\end{theorem}
\begin{proof}
Introduce registers $G_1, G_2$ such that $\ket{\psi}_{AA'A''G_1}$ and $\ket{\psi}_{BB'B''G_2}$ purify $\psi_{AA'A''}$ and $\psi_{BB'B''}$. Let $\rho_{CA'B'A''B''}\defeq \cN_{AB \to C}(\psi_{AA'A''}\otimes \psi_{BB'B''})$. Let $\ket{\kappa}_{E'A'}, \ket{\kappa}_{E''A''}, \ket{\kappa}_{F'B'}, \ket{\kappa}_{F''B''}$ be purifications of $\psi_{A'}, \psi_{A''}, \psi_{B'}, \psi_{B''}$. Let $V_A: \cH_{E''E'}\rightarrow \cH_{AG_1}$ be an isometry such that $V_A\ket{\kappa}_{E'A'}\otimes \ket{\kappa}_{E''A''} = \ket{\psi}_{AA'A''G_1}$ and $V_B: \cH_{F''F'}\rightarrow \cH_{BG_2}$ be an isometry such that $V_B\ket{\kappa}_{F'B'}\otimes \ket{\kappa}_{F''B''} = \ket{\psi}_{BB'B''G_2}$.

Alice and Charlie share one copy of $\ket{\kappa}_{E''A''}$, where Alice holds $E''$ and Charlie holds $A''$, and $2^{R_1}$ copies of $\ket{\kappa}_{E'A'}$ in registers $E'_1, E'_2, \ldots E'_{2^{R_1}}$, where Alice holds $E'_1, \ldots E'_{2^{R_1}}$ and Charlie holds $A'_1, \ldots A'_{2^{R_1}}$. Bob and Charlie share one copy of $\ket{\kappa}_{F''B''}$, where Bob holds $F''$ and Charlie holds $B''$, and $2^{R_2}$ copies of $\ket{\kappa}_{F'B'}$ in registers $F'_1B'_1, \ldots F'_{2^{R_2}}B'_{2^{R_2}}$, where Bob holds $F'_1, \ldots F'_{2^{R_2}}$ and Charlie holds $A'_1, \ldots A'_{2^{R_1}}$.

\vspace{0.1in}

\noindent {\bf Encoding:} To send the message $m_1 \in [2^{R_1}]$, Alice applies the isometry $V_A$ on registers $E''E'_{m_1}$ to prepare the purification $\ket{\psi}_{AA''A'_{m_1}G_1}$ of $\psi_{A''}\otimes \psi_{A'_{m_1}}$. She sends $A$ through the channel. 

To send the message $m_2$, Bob applies an isometry $V_B$ on the registers $F''F'_{m_2}$ to prepare the purification $\ket{\psi}_{BB''B'_{m_2}G_2}$ of $\psi_{B''}\otimes \psi_{B'_{m_2}}$. He sends $B$ through the channel.

\vspace{0.1in}

\noindent {\bf Decoding:} Let $\Pi_{CA''B''A'}$ be the operator achieving the optimum in $\dzeroseps{\eps_1}(\rho_{A'CA''B''}\| \rho_{CA''B''}\otimes \rho_{A'})$ and $\Pi_{CA''B''B'}$ be the operator achieving the optimum in $\dzeroseps{\eps_2}(\rho_{B'CA''B''}\| \rho_{CA''B''}\otimes \rho_{B'})$. By Stinespring dilation theorem, we introduce a register $J$ in the state $\ket{0}_J$ and consider the projectors $\Pi_{CA''B''A'J}$ and $\Pi_{CA''B''B'J}$. 

Charlie sequentially applies the measurement $\{\Pi_{CA''B''A'_{m_1}J}, \id - \Pi_{CA''B''A'_{m_1}J}\}$, for $m_1$ ranging from $[1: 2^{R_1}]$. He outputs the first $m_1$ for which he obtains the outcome corresponding to $\Pi_{CA''B''A'_{m_1}J}$. Then he sequentially applies the measurement $\{\Pi_{CA''B''B'_{m_2}J}, \id - \Pi_{CA''B''B'_{m_2}J}\}$, for $m_2$ ranging from $[1: 2^{R_2}]$. He outputs the first $m_2$ for which he obtains the outcome corresponding to $\Pi_{CA''B''B'_{m_2}J}$.

\vspace{0.1in}

\noindent {\bf Error analysis:} We compute the probability of obtaining the correct outcome. Let $\omega_{m_1, m_2}$ denote the overall quantum state with Charlie, conditioned on messages $m_1, m_2$. Let $\bar{\Pi}$ denote the projector orthogonal to $\Pi$. We have 
\begin{eqnarray*}
&&\Pr(M'_1, M'_2 = m_1, m_2\mid M_1, M_2 = m_1, m_2) \\ && = \Tr\bigg( \bar{\Pi}_{CA''B''B'_{2^{R_2}}J}\ldots \Pi_{CA''B''B'_{m_2}J}\ldots\bar{\Pi}_{CA''B''B'_{1}J}\bar{\Pi}_{CA''B''A'_{2^{R_1}}J}\ldots \Pi_{CA''B''A'_{m_1}J}\\ &&  \hspace{1cm}\ldots\bar{\Pi}_{CA''B''A'_{1}J} (\omega_{m_1, m_2}) \bar{\Pi}_{CA''B''A'_{1}J}\ldots\Pi_{CA''B''A'_{m_1}J}\ldots\bar{\Pi}_{CA''B''A'_{2^{R_1}}J}\bar{\Pi}_{CA''B''B'_{1}J}\\ && \hspace{1cm}\ldots\Pi_{CA''B''B'_{m_2}J}\ldots \bar{\Pi}_{CA''B''B'_{2^{R_2}}J} \bigg)\\
&& \geq 1 - 4\cdot\bigg(\Tr(\bar{\Pi}_{CA''B''A'_{m_1}J}\omega_{m_1, m_2}) + \Tr(\bar{\Pi}_{CA''B''B'_{m_2}J}\omega_{m_1,m_2}) \\ &&\hspace{1cm} + \sum_{m'_1\neq m_1}\Tr(\Pi_{CA''B''A'_{m'_1}J}\omega_{m_1, m_2}) + \sum_{m'_2\neq m_2}\Tr(\Pi_{CA''B''B'_{m'_2}J}\omega_{m_1, m_2})\bigg) \\
&& = 1 - 4\cdot\bigg(\Tr(\rho_{A'_{m_1}CA''B''}\otimes \ketbra{0}_J\bar{\Pi}_{CA''B''A'_{m_1}J}) + \Tr(\rho_{B'_{m_2}CA''B''}\otimes \ketbra{0}_J\bar{\Pi}_{CA''B''B'_{m_2}J}) \\ &&\hspace{1cm} + \sum_{m'_1\neq m_1}\Tr(\rho_{A'_{m'_1}}\otimes \rho_{CA''B''}\otimes \ketbra{0}_J\Pi_{CA''B''A'_{m'_1}J}) \\ &&\hspace{1cm} + \sum_{m'_2\neq m_2}\Tr(\rho_{B'_{m'_2}}\otimes \rho_{CA''B''}\otimes \ketbra{0}_J\Pi_{CA''B''B'_{m'_2}J})\bigg) \\
&& \geq 1- 4\cdot\bigg(\eps_1+\eps_2 + 2^{R_1 - \dzeroseps{\eps_1}(\rho_{A'CA''B''}\| \rho_{CA''B''}\otimes \rho_{A'})}+ 2^{R_2 - \dzeroseps{\eps_1}(\rho_{B'CA''B''}\| \rho_{CA''B''}\otimes \rho_{B'})}\bigg)\\
&& \geq 1- 4\cdot (\eps_1+\eps_2 + 2\delta),
\end{eqnarray*}
where in the first inequality, we use Fact \ref{noncommutativebound} and in last step, we use the bound on $R_1, R_2$. Thus, we conclude that 
$$\Pr(M'_1, M'_2 = m_1, m_2\mid M_1, M_2 = m_1, m_2) \leq 4\cdot (\eps_1+\eps_2 + 2\delta).$$
This completes the proof.
\end{proof}

\section{Entanglement unassisted quantum coding}
\label{sec:unassistqcode}

Similar bounds can be obtained for entanglement unassisted quantum coding. For brevity, we consider the average case error, although all of the achievability results below also hold worst case over the messages.

\subsection{Point to point quantum channel}

Alice wants to communicate a classical message $M$ chosen uniformly from $[2^R]$ to Bob over a quantum channel such that Bob is able to decode the correct message with probability at least $1-\eps$ , for all message $m$. Let the input to Alice be given in a register $M$. We now make the following definition.
\begin{definition}
An $(R, \eps )$ - code for the quantum channel $\cN_{A \to B}$ consists of 
\begin{itemize}
\item An encoding map $\cU : M \rightarrow A$ for Alice, where $M$ takes value uniformly over the set $[2^R]$.  
\item A decoding operation $\cD : B\rightarrow M'$ for Bob, with $M'\equiv M$ being the output register such that 
\beq
\Pr(M'\neq M) \leq \eps. \nonumber
\enq
\end{itemize}
\end{definition}

We have the following achievability and converse, obtaining results similar to that in \cite{WangR12}. In below, $\psi_{UA}$ is a classical-quantum state with $U$ being the classical register.

\begin{theorem}
\label{ptopconverse}
Fix a quantum channel $\cN_{A\to B}$ and $\eps\in (0,1)$. For any $(R, \eps)$- code for this quantum channel, it holds that 
$$R \leq \max_{\psi_{UA}: |U|\leq |B|^{\frac{1}{1-\eps}}, \psi_U= \frac{\id_U}{|U|}}\min_{\sigma_B}\dzeroseps{\eps}(\cN_{A\to B}(\psi_{UA})\|\sigma_B\otimes \psi_{U}).$$ Further, for every $\delta \in (0,1)$, there exists an $(R, \eps+\delta)$-code for this quantum channel, if  
$$R \leq \max_{\psi_{UA}: |U|\leq |B|^{\frac{1}{1-\eps}}}\dzeroseps{\eps}(\cN_{A\to B}(\psi_{UA})\|\cN_{A\to B}(\psi_A)\otimes \psi_{U}) - \log\frac{4\eps}{\delta^2}.$$
\end{theorem}
\begin{proof}
We first show the converse for uniform distribution over the message. Fix a quantum state $\sigma_B$. Let $\psi_{MA}$ be the quantum state after Alice's encoding and $\phi_{MM'}$ be the quantum state after Bob's decoding. From Facts \ref{fact:dhallstates} and \ref{fact:monotonequantumoperation} 
$$R\leq \dzeroseps{\eps}(\phi_{MM'}\| \phi_M\otimes \cD(\sigma_B)) \leq \dzeroseps{\eps}(\cN_{A\to B}(\psi_{MA})\| \psi_M\otimes \sigma_B).$$
Further, from Fact \ref{dhandd}, $$R\leq \dzeroseps{\eps}(\cN_{A\to B}(\psi_{MA})\| \psi_M\otimes \cN_{A\to B}(\psi_A)) \leq \frac{\relent{\cN_{A\to B}(\psi_{MA})}{\psi_M\otimes \cN_{A\to B}(\psi_A)}}{1-\eps} \leq \frac{\log|B|}{1-\eps}.$$
The converse now follows by setting $M=U$ and the fact that the state on register $M$ is uniform. 

The achievability is similar to the proof of Theorem \ref{theo:eaptopachieve}, where Alice and Bob share $2^R$ perfectly correlated copies of $\psi_U$ (as shared randomness). For sending $m\in [2^R]$, Alice inputs the register $A$ generated from $U_m$ according to the state $\psi_{UA}$. Bob performs the position-based decoding strategy to recover the message $m$. A protocol without randomness assistance is obtained since there exists a string $u_1, \ldots u_{2^R}$ for which the error probability is maintained.
\end{proof}

\subsection{Quantum channel with state}

Alice wants to communicate a classical message $M$ chosen from $[2^R]$ to Bob over a quantum channel $\cN_{AS \to B }$ such that Bob is able to decode the correct message with probability at least $1-\eps$. Alice shared entanglement $\ket{\tau}_{SS'}$ with the channel.
\begin{definition}
Let $\ket{\tau}_{SS'}$ be the state shared between Alice and channel. An $(R, \eps )$- code for the quantum channel $\cN_{ AS \to B}$ consists of 
\begin{itemize}
\item An encoding operation $\cE: MS' \rightarrow A$ for Alice, where $M$ takes values uniformly over $[2^R]$. 
\item A decoding operation $\cD : B\rightarrow M'$ for Bob, with $M'\equiv M$ being the output register such that
\begin{equation*}
\Pr(M'\neq M)\leq \eps. 
\end{equation*}
\end{itemize}
\end{definition}

We have the following theorem. Below, $\psi_{ASU}$ is a classical-quantum state with $U$ being the classical register. 

\begin{theorem}
Fix a quantum channel $\cN_{AS \to B}$ with state $\tau_S$ and an $\eps\in (0,1)$. For every $(R, \eps)$- code for this channel, it holds that
$$R \leq \max_{\psi_{ASU}: \psi_{SU} = \tau_S\otimes \frac{\id_U}{|U|}, |U|\leq |B|^{\frac{1}{1-\eps}}}\min_{\sigma_B}\dzeroseps{\eps}(\cN_{AS\to B}(\psi_{ASU})\|\psi_{U} \otimes \sigma_B).$$
Further for every $\delta\in (0,1)$, there exists an $(R, \eps + 2\delta)$- code for this channel, if  
$$R \leq \max_{\psi_{ASU}: \psi_{SU} = \tau_S\otimes \psi_U}\dzeroseps{\eps}(\cN_{AS\to B}(\psi_{ASU})\|\psi_{U} \otimes \cN_{AS\to B}(\psi_{AS})) - \log\frac{4\eps}{\delta^2}.$$
\end{theorem}
\begin{proof}
Let $\psi_{ASM}$ be the state after Alice's encoding. Fix a quantum state $\sigma_B$. Observe that $\psi_{MS}=\frac{\id_M}{|M|}\otimes \tau_S$. Let $\rho_{BM}$ be the state after the action of the channel and $\phi_{MM'}$ be the state after Bob's decoding. Then,
$$R\leq \dzeroseps{\eps}(\phi_{MM'}\|\phi_M\otimes \cD(\sigma_B))\leq \dzeroseps{\eps}(\rho_{MB}\|\rho_M \otimes \sigma_B).$$
Further, from Fact \ref{dhandd},
$$\log|M| = R\leq \dzeroseps{\eps}(\rho_{MB}\|\rho_M \otimes \rho_B) \leq \frac{\log|B|}{1-\eps}.$$
Setting $U=M$, we obtain the converse. 

The achievability follows similar to the proof of Theorem \ref{theo:qgelpinachieve}. Alice and Bob share $2^R$ perfectly correlated copies of the state $\psi_U$ (as shared randomness). To send the message $m$, Alice considers the register $U_m$ and applies an isometry on the register $S'$ of $\ket{\tau}_{SS'}$ to obtain the state $\psi_{ASU_m}$. She sends the register $A$ through the channel. Bob performs the position based decoding strategy to decode the message $m$. A protocol without randomness assistance is obtained since there exists a string $u_1, \ldots u_{2^R}$ for which the error probability is maintained.
\end{proof}

\subsection{Broadcast quantum channel}

Alice wishes to communicate message pair $(m_1,m_2)$ simultaneously to Bob and Charlie over a quantum broadcast channel, where $m_1$ is for Bob and $m_2$ is for Charlie, such that both Bob and Charlie output the correct message with probability at least $1-\eps$. 
\begin{definition}
An $(R_1, R_2, \eps_1, \eps_2)$ entanglement assisted code for the quantum broadcast channel $\cN_{ A \to BC}$ consists of 
\begin{itemize}
\item An encoding operation $\cE: M_1M_2 \rightarrow A$ for Alice, where $M_1, M_2$ take values uniformly over the sets $[2^{R_1}], [2^{R_2}]$ respectively.
\item A pair of decoding operations $(\cD_B,\cD_C)$,  $\cD_B : B\rightarrow M'_1$ and $\cD_C : C \rightarrow M'_2$, with $(M_1',M_2')\equiv (M_1,M_2)$ being the output registers, such that 
\begin{equation*}
\Pr(M'_1\neq M_1)\leq \eps_1, \Pr(M'_2\neq M_2)\leq \eps_2.
\end{equation*}
\end{itemize}
\end{definition}

We have the following theorem.

\begin{theorem}
Fix a quantum channel $\cN_{A\to BC}$ and $\eps_1, \eps_2 \in (0,1)$. For any $(R_1, R_2, \eps_1, \eps_2)$- code for this channel, there exist registers $U, V$ such that $|U||V|\leq (|B||C|)^{\frac{1}{1-\eps_1-\eps_2}}$ and a classical-quantum state $\psi_{AUV}$ satisfying $\psi_{UV}= \frac{\id_U}{|U|}\otimes \frac{\id_V}{|V|}$, with registers $U,V$ classical, such that
$$R_1 \leq \min_{\sigma_B}\dzeroseps{\eps_1}(\Tr_C\cN_{A\to BC}(\psi_{AU})\|\sigma_B \otimes \psi_{U}),$$
$$R_2 \leq \min_{\tau_C}\dzeroseps{\eps_2}(\Tr_B\cN_{A\to BC}(\psi_{AV})\|\tau_C \otimes \psi_{V}).$$ 
Furthermore, for every $\delta\in (0,1)$ and classical-quantum state $\psi_{AUV}$ satisfying $\psi_{UV}= \psi_U\otimes \psi_V$, with registers $U,V$ being classical, there exists an $(R_1, R_2, \eps_1 + 2\delta, \eps_2 + 2\delta)$- code for this channel, if 
$$R_1 \leq \dzeroseps{\eps_1}(\Tr_C\cN_{A\to BC}(\psi_{AU})\|\Tr_C\cN_{A\to BC}(\psi_A) \otimes \psi_{U}) - \log\frac{4\eps_1}{\delta^2},$$
$$R_2 \leq \min\dzeroseps{\eps_2}(\Tr_B\cN_{A\to BC}(\psi_{AV})\|\Tr_B\cN_{A\to BC}(\psi_A) \otimes \psi_{V})- \log\frac{4\eps_2}{\delta^2}.$$ 
\end{theorem}
\begin{proof}
We first show the converse for uniform distribution over messages. Fix quantum states $\sigma_B, \tau_C$. Let $\psi_{AM_1M_2}$ be the quantum state after Alice's encoding, $\rho_{BCM_1M_2}$ be the quantum state after the action of the channel and $\phi_{M'_1M'_2M_1M_2}$ be the quantum state after Bob's and Charlie's decoding. Observe that $\psi_{M_1M_2} = \phi_{M_1M_2} = \frac{\id_{M_1}}{|M_1|}\otimes \frac{\id_{M_2}}{|M_2|}$. From Facts \ref{fact:dhallstates} and \ref{fact:monotonequantumoperation},
$$R_1\leq \dzeroseps{\eps_1}(\phi_{M_1M'_1}\| \phi_{M_1}\otimes \cD_B(\sigma_B)) \leq \dzeroseps{\eps_1}(\rho_{M_1B}\| \rho_{M_1}\otimes \sigma_B).$$ Similarly, 
$$R_2\leq \dzeroseps{\eps_2}(\rho_{M_2C}\| \rho_{M_2}\otimes \tau_C).$$
Finally, from Fact \ref{dhandd},
\begin{eqnarray*}
\log|M_1||M_2| &=& R_1+R_2 \leq \dzeroseps{\eps_1+\eps_2}(\phi_{M_1M_2M'_1M'_2}\|  \phi_{M_1M_2}\otimes \phi_{M'_1M'_2}) \leq \dzeroseps{\eps_1+\eps_2}(\psi_{M_1M_2A}\| \psi_{M_1M_2}\otimes \psi_A)\\
&\leq& \frac{\relent{\psi_{M_1M_2BC}}{\psi_{M_1M_2}\otimes \psi_{BC}}}{1-\eps_1-\eps_2} \leq \frac{\log|B||C|}{1-\eps_1-\eps_2}.
\end{eqnarray*}
Setting $U= M_1, V= M_2$, the converse follows. 

The achievability follows similar to the proof of Theorem \ref{theo:qbroadachieve}. Alice and Bob share $2^{R_1}$ perfectly correlated copies of the state $\psi_U$ (as shared randomness). Alice and Charlie share $2^{R_2}$ perfectly correlated copies of the state $\psi_V$ (as shared randomness). To send the messages $(m_1, m_2)$, Alice inputs the register $A$ obtained from $U_{m_1}, V_{m_2}$ according to the quantum state $\psi_{AU_{m_1}V_{m_2}}$. Bob and Charlie respectively perform the position  based decoding strategy to obtain the messages $m_1, m_2$. A protocol without randomness assistance is obtained since there exists a string $u_1, \ldots u_{2^{R_1}}, v_1, \ldots v_{2^{R_2}}$ for which the error probability is maintained.
\end{proof}

\subsection{Multiple-access channel}

Alice wants to communicate a classical message $m_1$ chosen uniformly from $[2^{R_1}]$ to Charlie and Bob wants to communicate a classical message $m_2$ chosen uniformly from $[2^{R_2}]$ to Charlie, over a channel $\cN_{AB \to C}$. This is the multiple access channel. 
\begin{definition}
An $(R_1, R_2, \eps_1, \eps_2)$- code for the quantum multiple access channel $\cN_{AB \to C}$ consists of 
\begin{itemize}
\item Encoding operations $\cE_1: M_1 \rightarrow A$ and $\cE_2: M_2\rightarrow B$, where $M_1, M_2$ take values uniformly over the sets $[2^{R_1}], [2^{R_2}]$ respectively. 
\item A decoding operation $\cD: C \rightarrow M'_1M'_2$, with $M'_1\equiv M_1$ and $M'_2\equiv M_2$ such that for all $(m_1, m_2)$,
$$\Pr(M'_1\neq M_1|)\leq \eps_1, \Pr(M'_2\neq M_2)\leq \eps_2.$$ 
\end{itemize}  
\end{definition}

We have the following result.

\begin{theorem}
Fix a quantum channel $\cN_{AB\to C}$ and $\eps_1, \eps_2\in (0,1)$. For every $(R_1, R_2, \eps_1, \eps_2)$- code for this channel, there exist registers $U, V'$ satisfying $|U'||V'| \leq |C|^{\frac{1}{1-\eps_1-\eps_2}}$ and classical-quantum states $\psi_{UA}, \psi_{VB}$ satisfying $\psi_U = \frac{\id_U}{|U|}$ and $\psi_{V'} = \frac{\id_V}{|V|}$ such that 
$$R_1 \leq \min_{\sigma_C}\dzeroseps{\eps_1}(\rho_{CU}\| \sigma_{C}\otimes \rho_{U}),$$
and
$$R_2 \leq \min_{\tau_C}\dzeroseps{\eps_2}(\rho_{CV}\| \tau_{C}\otimes \rho_{V}),$$
where $\rho_{CUV} \defeq \cN_{AB\to C}(\psi_{UA}\otimes \psi_{VB})$.

Furthermore, for every classical-quantum state $\psi_{UA}, \psi_{VB}$, there exists a $(R_1, R_2, 4(\eps_1+\eps_2 + 2\delta), 4(\eps_1+\eps_2 + 2\delta))$- code for this channel, if
$$R_1 \leq \dzeroseps{\eps_1}(\rho_{CU}\| \rho_{C}\otimes \rho_{U}) - \log\frac{1}{\delta},$$
and
$$R_2 \leq \dzeroseps{\eps_2}(\rho_{CV}\| \rho_{C}\otimes \rho_{V}) - \log\frac{1}{\delta},$$
where $\rho_{CUV} \defeq \cN_{AB\to C}(\psi_{UA}\otimes \psi_{VB})$. In fact, the following upper bound holds, 
$$\Pr(M'_1, M'_2 \neq M_1, M_2 ) \leq 4\cdot (\eps_1+\eps_2 + 2\delta).$$
\end{theorem}

\begin{proof}
We first show the converse. Fix quantum states $\sigma_C, \tau_C$. Let $\psi_{M_1A}\otimes \psi_{M_2B}$ be the quantum state after Alice's encoding and $\rho_{M_1M_2C}$ be the state after the action of the channel. Let $\phi_{M_1M_2M'_1M'_2}$ be the state after Charlie's decoding. From Facts \ref{fact:dhallstates} and \ref{fact:monotonequantumoperation},
\begin{eqnarray*}
R_1 &\leq& \dzeroseps{\eps_1}(\phi_{M_1M'_1M'_2}\|\phi_{M_1}\otimes \cD(\sigma_C))\leq \dzeroseps{\eps_1}(\rho_{M_1C}\|\rho_{M_1}\otimes \sigma_C).
\end{eqnarray*}
Similarly, 
$$R_2 \leq \dzeroseps{\eps_1}(\rho_{M_2C}\|\rho_{M_2}\otimes \tau_C).$$
Further, from Fact \ref{dhandd},
$$\log|M_1||M_2| = R_1+R_2\leq \dzeroseps{\eps_1}(\rho_{M_1M_2C}\|\rho_{M_1M_2}\otimes \rho_C)\leq \frac{\log|C|}{1-\eps_1-\eps_2}.$$
We set $U=M_1, V=M_2$, which proves the converse. 

The achievability is similar to the proof of Theorem \ref{qmultaccessgood}. Alice and Charlie share $2^{R_1}$ perfectly correlated copies of the state $\psi_U$ (as shared randomness). Bob and Charlie share $2^{R_2}$ perfectly correlated copies of $\psi_V$ (as shared randomness). To send message $m_1$, Alice inputs the register $A$ generated from $U_{m_1}$ according to the quantum state $\psi_{AU_{m_1}}$. To send message $m_2$, Bob inputs the register $B$ generated from $V_{m_2}$ according to the quantum state $\psi_{BV_{m_1}}$. Charlie performs the sequential position-based decoding, decoding message $m_1$ and then $m_2$. Another protocol is obtained where Charlie decodes $m_2$ and then $m_1$. The error analysis follows in a similar fashion. A protocol without randomness assistance is obtained since there exists a string $u_1, \ldots u_{2^{R_1}}, v_1, \ldots v_{2^{R_2}}$ for which the error probability is maintained.
\end{proof}

\subsection*{Conclusion}

We have obtained a near-optimal one-shot characterization of the amount of communication for a wide family of quantum channels in the one-shot setting. Our one-shot bounds for the entanglement-unassisted case (Section \ref{sec:unassistqcode}) have the property that the register sizes involved in the bounds are bounded. We leave the task of obtaining similar near-optimal bounds for entanglement-assisted cases with bounded register dimensions (except for the point to point case) in Section \ref{sec:entqcode} for future work.

We stress that similar results could also be obtained for the classical case. But this would lead to formulas that are not single letter in the asymptotic and i.i.d. setting \cite[Section 4.3]{GamalK12} (except for the point to point classical channel). Interestingly, there are alternative characterizations known for the multiple access classical channel and classical channel with state, which lead to single letter optimal bounds in the asymptotic and i.i.d. setting and near optimal bounds for the one-shot setting. Unfortunately, as discussed earlier, analogous bounds are not known to be single letter for the quantum multiple access channel (nor are they known to be near-optimal in the one-shot setting). Finding a single letter optimal rate region is an important open question in the asymptotic and i.i.d. theory of quantum channel coding over networks.

\subsection*{Acknowledgment} 

This work is supported by the Singapore Ministry of Education and the National Research Foundation,
also through the Tier 3 Grant “Random numbers from quantum processes” MOE2012-T3-1-009 and NRF RF
Award NRF-NRFF2013-13.

\bibliographystyle{ieeetr}
\bibliography{References}

\begin{thebibliography}{10}

\bibitem{AnshuJW17CC}
A.~Anshu, R.~Jain, and N.~Warsi, ``One-shot entanglement assisted classical and
  quantum communication over noisy quantum channels: A hypothesis testing and
  convex-split approach.'' https://arxiv.org/abs/1702.01940, 2017.

\bibitem{MatthewsW14}
W.~Matthews and S.~Wehner, ``Finite blocklength converse bounds for quantum
  channels,'' {\em IEEE Transactions on Information Theory}, vol.~60,
  pp.~7317--7329, Nov 2014.

\bibitem{Shannon}
C.~E. Shannon, ``A mathematical theory of communication,'' {\em The Bell System
  Technical Journal}, vol.~27, pp.~379--423, July 1948.

\bibitem{Ahlswede73}
R.~Ahlswede, {\em {Multi-way communication channels}}, pp.~23 -- 51.
\newblock Akadémiai Kiadó, 1973.

\bibitem{Liao72}
H.~Liao, ``Multiple access channels,'' 1972.
\newblock PhD Thesis, Department of Electrical Engineering, University of
  Hawaii, Honolulu.

\bibitem{GelfandP80}
S.~I. Gelfand and M.~S. Pinsker, ``Coding for channels with random
  parameters,'' {\em Problem of Control and Information Theory}, vol.~9, no.~1,
  pp.~19--31, 1980.

\bibitem{Marton79}
K.~Marton, ``A coding theorem for the discrete memoryless broadcast channel,''
  {\em IEEE Transactions on Information Theory}, vol.~25, pp.~306--311, May
  1979.

\bibitem{GamalK12}
A.~E. Gamal and Y.-H. Kim, {\em Network Information Theory}.
\newblock New York, NY, USA: Cambridge University Press, 2012.

\bibitem{BennettSST02}
C.~H. Bennett, P.~W. Shor, J.~A. Smolin, and A.~V. Thapliyal,
  ``Entanglement-assisted capacity of a quantum channel and the reverse shannon
  theorem,'' {\em IEEE Transactions on Information Theory}, vol.~48,
  pp.~2637--2655, Oct 2002.

\bibitem{Holevo98}
A.~S. Holevo, ``The capacity of the quantum channel with general signal
  states,'' {\em IEEE Transactions on Information Theory}, vol.~44,
  pp.~269--273, Jan 1998.

\bibitem{SchuW97}
B.~Schumacher and M.~D. Westmoreland, ``Sending classical information via noisy
  quantum channels,'' {\em Phys. Rev. A}, vol.~56, pp.~131--138, Jul 1997.

\bibitem{lloyd97}
S.~Lloyd, ``Capacity of the noisy quantum channel,'' {\em Phys. Rev. A},
  vol.~55, pp.~1613--1622, Mar 1997.

\bibitem{Shor02}
P.~Shor, ``The quantum channel capacity and coherent information.'' Lecture
  Notes, MSRI Workshop on Quantum Computation., 2002.

\bibitem{Devetak05private}
I.~Devetak, ``The private classical capacity and quantum capacity of a quantum
  channel,'' {\em IEEE Transactions on Information Theory}, vol.~51,
  pp.~44--55, Jan 2005.

\bibitem{Teleportation93}
C.~H. Bennett, G.~Brassard, C.~Cr\'epeau, R.~Jozsa, A.~Peres, and W.~K.
  Wootters, ``Teleporting an unknown quantum state via dual classical and
  einstein-podolsky-rosen channels,'' {\em Phys. Rev. Lett.}, vol.~70,
  pp.~1895--1899, Mar 1993.

\bibitem{bennett92}
C.~H. Bennett and S.~J. Wiesner, ``Communication via one- and two-particle
  operators on einstein-podolsky-rosen states,'' {\em Phys. Rev. Lett.},
  vol.~69, no.~20, pp.~2881--2884, 1992.

\bibitem{Winter01}
A.~Winter, ``The capacity of the quantum multiple-access channel,'' {\em IEEE
  Transactions on Information Theory}, vol.~47, pp.~3059--3065, Nov 2001.

\bibitem{HsiehDW08}
M.~H. Hsieh, I.~Devetak, and A.~Winter, ``Entanglement-assisted capacity of
  quantum multiple-access channels,'' {\em IEEE Transactions on Information
  Theory}, vol.~54, pp.~3078--3090, July 2008.

\bibitem{YardHD08}
J.~Yard, P.~Hayden, and I.~Devetak, ``Capacity theorems for quantum
  multiple-access channels: classical-quantum and quantum-quantum capacity
  regions,'' {\em IEEE Transactions on Information Theory}, vol.~54,
  pp.~3091--3113, July 2008.

\bibitem{FawziHSSW12}
O.~Fawzi, P.~Hayden, I.~Savov, P.~Sen, and M.~M. Wilde, ``Classical
  communication over a quantum interference channel,'' {\em IEEE Transactions
  on Information Theory}, vol.~58, pp.~3670--3691, June 2012.

\bibitem{XuW13}
S.~C. Xu and M.~M. Wilde, ``Sequential, successive, and simultaneous decoders
  for entanglement-assisted classical communication,'' {\em Quantum Information
  Processing}, vol.~12, pp.~641--683, Jan 2013.

\bibitem{SaakianA98}
A.~Allahverdyan and D.~Saakian, ``The broadcast quantum channel for classical
  information transmission.'' https://arxiv.org/abs/quant-ph/9805067, 1998.

\bibitem{DupuisHL10}
F.~Dupuis, P.~Hayden, and K.~Li, ``A father protocol for quantum broadcast
  channels,'' {\em IEEE Transactions on Information Theory}, vol.~56,
  pp.~2946--2956, June 2010.

\bibitem{YardHD11}
J.~Yard, P.~Hayden, and I.~Devetak, ``Quantum broadcast channels,'' {\em IEEE
  Transactions on Information Theory}, vol.~57, pp.~7147--7162, Oct 2011.

\bibitem{SavovW15}
I.~Savov and M.~M. Wilde, ``Classical codes for quantum broadcast channels,''
  {\em IEEE Transactions on Information Theory}, vol.~61, pp.~7017--7028, Dec
  2015.

\bibitem{Dupuis09}
F.~Dupuis, ``The capacity of quantum channels with side information at the
  transmitter,'' in {\em 2009 IEEE International Symposium on Information
  Theory}, pp.~948--952, June 2009.

\bibitem{Frederic10}
F.~Dupuis, ``The decoupling approach to quantum information theory,'' 2010.
\newblock PhD Thesis, Universit{\'e} de Montr{\'e}al.

\bibitem{BuscemiD10}
F.~Buscemi and N.~Datta, ``The quantum capacity of channels with arbitrarily
  correlated noise,'' {\em IEEE Transactions on Information Theory}, vol.~56,
  pp.~1447--1460, 2010.

\bibitem{WangR12}
L.~Wang and R.~Renner, ``One-shot classical-quantum capacity and hypothesis
  testing,'' {\em Phys. Rev. Lett.}, vol.~108, p.~200501, May 2012.

\bibitem{DattaH13}
N.~Datta and M.~H. Hsieh, ``One-shot entanglement-assisted quantum and
  classical communication,'' {\em IEEE Transactions on Information Theory},
  vol.~59, pp.~1929--1939, March 2013.

\bibitem{DattaTW2016}
N.~Datta, M.~Tomamichel, and M.~M. Wilde, ``On the second-order asymptotics for
  entanglement-assisted communication,'' {\em Quantum Information Processing},
  vol.~15, no.~6, pp.~2569--2591, 2016.

\bibitem{WangFT17}
X.~Wang, K.~Fang, and M.~Tomamichel, ``On converse bounds for classical
  communication over quantum channels.'' https://arxiv.org/abs/1709.05258,
  2017.

\bibitem{XieWD17}
W.~Xie, X.~Wang, and R.~Duan, ``Converse bounds for classical communication
  over quantum networks.'' https://arxiv.org/abs/1712.05637, 2017.

\bibitem{AnshuDJ14}
A.~Anshu, V.~K. Devabathini, and R.~Jain, ``Quantum communication using
  coherent rejection sampling,'' {\em Phys. Rev. Lett.}, vol.~119, p.~120506,
  Sep 2017.

\bibitem{Wilde17}
M.~M. Wilde, {\em Quantum Information Theory}.
\newblock Cambridge University Press, 2~ed., 2017.

\bibitem{NayakS02}
A.~Nayak and J.~Salzman, ``On communication over an entanglement-assisted
  quantum channel,'' in {\em Proceedings of the Thiry-fourth Annual ACM
  Symposium on Theory of Computing}, STOC '02, (New York, NY, USA),
  pp.~698--704, ACM, 2002.

\bibitem{HayashiN03}
M.~Hayashi and H.~Nagaoka, ``General formulas for capacity of classical-quantum
  channels,'' {\em IEEE Transactions on Information Theory}, vol.~49,
  pp.~1753--1768, July 2003.

\bibitem{Sen12}
P.~Sen, ``Achieving the han-kobayashi inner bound for the quantum interference
  channel,'' in {\em 2012 IEEE International Symposium on Information Theory
  Proceedings}, pp.~736--740, July 2012.

\bibitem{Gao15}
J.~Gao, ``Quantum union bounds for sequential projective measurements,'' {\em
  Phys. Rev. A}, vol.~92, p.~052331, Nov 2015.

\bibitem{GioLM12}
V.~Giovannetti, S.~Lloyd, and L.~Maccone, ``Achieving the holevo bound via
  sequential measurements,'' {\em Phys. Rev. A}, vol.~85, p.~012302, Jan 2012.

\bibitem{Wilde13}
M.~M. Wilde, ``Sequential decoding of a general classical-quantum channel,''
  {\em Proceedings of the Royal Society of London A: Mathematical, Physical and
  Engineering Sciences}, vol.~469, no.~2157, 2013.

\bibitem{OMW18}
S.~K. Oskouei, S.~Mancini, and M.~M. Wilde, ``Union bound for quantum
  information processing.'' https://arxiv.org/abs/1804.08144, 2018.

\bibitem{DrescherF13}
L.~Drescher and O.~Fawzi, ``On simultaneous min-entropy smoothing,'' in {\em
  2013 IEEE International Symposium on Information Theory}, pp.~161--165, July
  2013.

\bibitem{QiWW17}
H.~Qi, Q.~Wang, and M.~M. Wilde, ``Applications of position-based coding to
  classical communication over quantum channels.''
  https://arxiv.org/abs/1704.01361, 2017.

\bibitem{Josza94}
R.~Jozsa, ``Fidelity for mixed quantum states,'' {\em Journal of Modern
  Optics}, vol.~41, no.~12, pp.~2315--2323, 1994.

\bibitem{uhlmann76}
A.~Uhlmann, ``The "transition probability" in the state space of a *-algebra,''
  {\em Rep. Math. Phys.}, vol.~9, pp.~273--279, 1976.

\bibitem{GilchristLN05}
A.~Gilchrist, N.~K. Langford, and M.~A. Nielsen, ``Distance measures to compare
  real and ideal quantum processes,'' {\em Phys. Rev. A}, vol.~71, p.~062310,
  Jun 2005.

\bibitem{umegaki1954}
H.~Umegaki, ``Conditional expectation in an operator algebra, i,'' {\em Tohoku
  Math. J. (2)}, vol.~6, no.~2-3, pp.~177--181, 1954.

\bibitem{Datta09}
N.~Datta, ``Min- and max- relative entropies and a new entanglement monotone,''
  {\em IEEE Transactions on Information Theory}, vol.~55, pp.~2816--2826, 2009.

\bibitem{Tomamichel12}
M.~Tomamichel, ``A framework for non-asymptotic quantum information theory,''
  2012.
\newblock PhD Thesis, ETH Zurich.

\bibitem{barnum96}
H.~Barnum, C.~M. Cave, C.~A. Fuch, R.~Jozsa, and B.~Schmacher, ``Noncommuting
  mixed states cannot be broadcast,'' {\em Phys. Rev. Lett.}, vol.~76, no.~15,
  pp.~2818--2821, 1996.

\bibitem{lindblad75}
G.~Lindblad, ``Completely positive maps and entropy inequalities,'' {\em
  Commun. Math. Phys.}, vol.~40, pp.~147--151, 1975.

\bibitem{Winter:1999}
A.~Winter, ``Coding theorem and strong converse for quantum channels.,'' {\em
  IEEE Transactions on Information Theory}, vol.~45, no.~7, pp.~2481--2485,
  1999.

\bibitem{Ogawa:2002}
T.~Ogawa and H.~Nagaoka, ``A new proof of the channel coding theorem via
  hypothesis testing in quantum information theory,'' in {\em Information
  Theory, 2002. Proceedings. 2002 IEEE International Symposium on}, pp.~73--,
  2002.

\bibitem{Watrouslecturenote}
J.~Watrous, ``Theory of {Q}uantum {I}nformation, lecture notes,'' 2011.
\newblock https://cs.uwaterloo.ca/~watrous/LectureNotes.html.

\end{thebibliography}

\end{document}